\documentstyle[]{mn}

\title[Six detached white-dwarf close binaries]{Six detached
  white-dwarf close binaries\thanks{The Isaac Newton and William
  Herschel telescopes are operated on the island of La Palma by the
  Isaac Newton Group in the Spanish Observatorio del Roque de los
  Muchachos of the Instituto de Astrof\'{\i}sica de Canarias. Based on
  observations collected at the Centro Astron\'{o}mico Hispano
  Alem\'{a}n (CAHA) at Calar Alto, operated jointly by the Max-Planck
  Institut f\"{u}r Astronomie and the Instituto de Astrof\'{\i}sica de
  Andaluc\'{\i}a (CSIC). Based on observations collected with ESO
  telescopes at the Paranal Observatory under programme IDs
  165.H.-0588 and 167.D-0407.}}

\author[L.\, Morales-Rueda et al.] {L.\,Morales-Rueda$^{1,2}$,
  T.\,R.\,Marsh$^{3,2}$, P.\,F.\,L.\,Maxted$^{4,2}$, G.\,Nelemans$^{5,1}$
  C.\,Karl$^{6}$ \newauthor R.\,Napiwotzki$^{7}$, C.\,K.\,J.\,Moran$^{2}$\\
  $^{1}$Department of Astrophysics, University of Nijmegen, P.O. Box
  9010, 6500 GL Nijmegen, The Netherlands (lmr@astro.kun.nl)\\
  $^{2}$Department of Physics and Astronomy, University of
  Southampton, Southampton SO17 1BJ, UK\\
  $^{3}$Department of Physics, University of Warwick, Coventry, CV4
  7AL, UK (T.R.Marsh@warwick.ac.uk)\\
  $^{4}$School of Chemistry and Physics, Keele University,
  Staffordshire ST5 5BG, UK (pflm@astro.keele.ac.uk)\\
  $^{5}$ Institute of Astronomy, Madingley Rd, University of
  Cambridge, Cambridge CB3 0HA, UK\\
  $^{6}$Dr Remeis-Sternwarte, Astronomisches Institut der Universit{\"
  a}t Erlangen-N{\" u}rnberg, Sternwarstrasse 7, 96049 Bamberg,
  Germany\\
  $^{7}$Department of Physics \& Astronomy, University of Leicester,
  University Road, Leicester LE1 7RH, UK
  }

\date{Accepted 0000 000 00; Received 0000 000 00; in original
  form 0000 000 00}

\pagerange{\pageref{firstpage}--\pageref{lastpage}}

\def\LaTeX{L\kern-.36em\raise.3ex\hbox{a}\kern-.15em
    T\kern-.1667em\lower.7ex\hbox{E}\kern-.125emX}

\begin{document}

\newcommand{\etal}{\mbox{et\ al.}}
\newcommand{\kmsec}{\,\mbox{$\mbox{km}\,\mbox{s}^{-1}$}}
\newcommand{\phiorb}{$\phi_{\tiny orb}$}
\newcommand{\msun}{\hbox{$\hbox{M}_\odot$}}
\newcommand{\rsun}{\hbox{$\hbox{R}_\odot$}}
\newcommand{\ha}{\hbox{$\hbox{H}\alpha$}}
\newcommand{\hb}{\hbox{$\hbox{H}\beta$}}
\newcommand{\hg}{\hbox{$\hbox{H}\gamma$}}
\newcommand{\heii}{\hbox{$\hbox{He\,{\sc ii}\,$\lambda$4686\,\AA}$}}
\newcommand{\hei}{\hbox{$\hbox{He\,{\sc i}\,$\lambda$4472\,\AA}$}}
\newcommand{\dda}{\mbox{WD1022+050}}
\newcommand{\ddb}{\mbox{WD1428+373}}
\newcommand{\ddc}{\mbox{WD1824+040}}
\newcommand{\ddd}{\mbox{WD2032+188}}
\newcommand{\dma}{\mbox{WD1042$-$690}}
\newcommand{\dmb}{\mbox{WD2009+622}}
\newcommand{\app}{\hbox{\AA\,pixel$^{-1}$}}

\label{firstpage}

\maketitle

\begin{abstract}
  
  We determine the orbits of four double degenerate systems (DDs),
  composed of two white dwarfs, and of two white dwarf -- M dwarf
  binaries. The four DDs, \dda, \ddb, \ddc, and \ddd, show orbital
  periods of 1.157155(5) d, 1.15674(2) d, 6.26602(6) d and 5.0846(3) d
  respectively. These periods combined with estimates for the masses
  of the brighter component, based on their effective temperatures,
  allow us to constrain the masses of the unseen companions. We
  estimate that the upper limit for the contribution of the unseen
  companions to the total luminosity in the four DDs ranges between 10
  and 20 per cent. In the case of the two white dwarf - M dwarf
  binaries, \dma\ and \dmb, we calculate the orbital parameters by
  fitting simultaneously the absorption line from the white dwarf and
  the emission core from the M-dwarf. Their orbital periods are
  0.337083(1) d and 0.741226(2) d respectively. We find signatures of
  irradiation on the inner face of \dmb's companion. We calculate the
  masses of both components from the gravitational redshift and the
  mass-radius relationship for white dwarfs and find masses of 0.75 --
  0.78 \msun\ and 0.61 -- 0.64 \msun\ for \dma\ and \dmb\
  respectively. This indicates that the stars probably reached the
  asymptotic giant branch in their evolution before entering a common
  envelope phase. These two white dwarf - M dwarf binaries will become
  cataclysmic variables, although not within a Hubble time, with
  orbital periods below the period gap.

\end{abstract}

\begin{keywords}
  
binaries: close -- binaries: spectroscopic -- white dwarfs

\end{keywords}

\section{Introduction}

About 10 per cent of white dwarfs reside in close binary systems
\cite{n04}. The formation of white-dwarf close-binary systems
commences with a pair of main-sequence stars orbiting in a wide
binary. The most massive star will evolve faster becoming a giant and
transferring mass to its companion. If mass transfer is sufficiently
rapid, the result will be the formation of a common envelope (CE) made
of the outer layers of the giant. The orbital energy of the binary can
then be used to eject the envelope, the result being a binary composed
of a white dwarf and a main sequence companion. If the mass ratio of
the initial components was close to unity, the binary formed after CE
ejection can still be wide, whereas in the case of extreme mass
ratios, the two components will spiral in to eject the envelope
forming a tighter binary. This binary, called a post common envelope
binary (PCEB), if close enough, might become a cataclysmic variable
(CV) if mass is transferred stably from the main sequence star to the
white dwarf. If, on the other hand, the binary is too wide to become a
CV, the main sequence star will evolve into a giant and the system
will undergo another CE phase. When the envelope is ejected, the
resulting binary will be composed of two white dwarfs a few solar
radii apart - called a double degenerate (DD). For details on the
evolution and formation of DDs see Nelemans et al.'s
\shortcite{nypv01} recent population synthesis studies and references
therein. The study of white-dwarf close-binary stars can help us
understand the elusive CE phase that they must have gone through, at
least once, during their evolution. This phase is very difficult to
study in any other manner as it lasts only of the order of 1 to 100
years. For recent calculations on CE ejection efficiencies see Soker
\&\ Harpaz \shortcite{sh03} and references therein.

The subjects of this paper, white dwarf - M dwarf binaries and DDs,
are, as mentioned above, the progenitors of CVs, and potential
progenitors of Type Ia supernova respectively. A short period DD where
the combined mass of the white dwarfs exceeds the Chandrasekhar mass
might become a Type Ia supernova \cite{it84}, but no candidates have
yet been found. We should mention that this idea is still quite
controversial with some theorists claiming that a DD does not become a
Type Ia supernova \cite{sn98} and others claiming that this can be the
case if one includes rotation in the calculations \cite{p03}. The ESO
Supernova Ia Progenitor Survey (SPY) has as its main goal to search
systematically for massive, short period DDs in the galaxy to
establish their direct link with Type Ia supernova and it is the
source of more than 100 recently discovered DD systems
\cite{n04}. Efforts are also been carried out to systematically find
white - M dwarf binaries by using the Sloan Digital Sky Survey (SDSS)
resulting in more than 400 possible candidates \cite{shs03}.

Schreiber \&\ G\"{a}nsicke \shortcite{sg03} review the evolution of
the known sample of white-dwarf close binaries with main sequence
companions, the PCEBs, and conclude that, due to selection effects,
the (until then) known population of PCEBs is dominated by young
systems with hot white dwarfs that will evolve into short period CVs
(P $<$ 3 h). The PCEB sample is biased toward low mass companions as
well which is the reason why they evolve into short period CVs. In
contrast, the PCEBs found in the SDSS \cite{r03}, published later on,
show an average white dwarf temperature significantly lower,
demonstrating that to sample the full parameter space better selection
criteria have to be devised.

In this paper we study 6 white dwarf close-binary stars. Four of them
are found to be DDs, the other two are composed of a white dwarf and
an M dwarf star. The separation of both components in these binaries
is of the order of a few solar radii, as determined from their short
orbital periods, so they are detached systems where no mass transfer
between them takes place. We obtain their orbital solutions and
compare the results obtained with predictions drawn from population
synthesis studies.

\section{Observations and Reduction}

The data used in this study were taken over many years (since 1993)
using five different telescopes and seven different
setups. Table~\ref{obs} gives a list of the number of spectra taken
for each target at each observing campaign indicating also which
telescope was used in each case.

\begin{table}
\caption{Journal of observations. A description of the setup used with
  each telescope is given in the text. N indicates the number of
  spectra taken with each setup. $^*$ indicates high resolution
  spectra covering only the emission core of \ha. The brightness in B
  magnitudes of each target is also given.}
\label{obs}
\begin{tabular}{lllll}
Object & N & Date & Setup & Inst.\\
\hline
\dda        & 10 & 29/2-3/3/96 & AAT & RGO\\
(B = 14.37) & 3  & 19/3/97 & AAT & RGO\\
            & 6  & 8-10/2/98 & INTa & IDS\\
            & 2  & 5/6/98 & AAT & RGO\\
            & 2  & 15/4/03 & INTb & IDS \\
            & 2  & 17/4/03 & INTb & IDS \\
\hline
\ddb       & 2  & 16/2/97 & INTb  & IDS\\
(B = 14.9) & 4  & 9-10/2/98 & INTa & IDS\\
           & 9  & 3-5/3/99  & INTa & IDS\\
           & 8  & 23/7/00   & WHT & ISIS\\
           & 2  & 23/1/03   & WHT & ISIS\\
\hline
\ddc        & 13 & 19-24/6/95 & INTa & IDS\\
(B = 14.00) & 13 & 29/2-3/3/96 & AAT & RGO\\
            & 17 & 18-21/3/97 & AAT & RGO\\
            & 2  & 23-24/6/97 & INTa & IDS\\
            & 2  & 17-18/5/00 & VLT & UVES\\
            & 5  & 4/5-18/8/01 & VLT & UVES\\
            & 10 & 26-30/10/01 & INTa & IDS\\
            & 5  & 23-27/2/02 & 3.5m & TWIN\\
\hline
\ddd        & 10 & 12-24/6/93 & WHT & ISIS\\
(B = 15.30) & 3  & 14-15/8/93 & WHT & ISIS\\
            & 8  & 10-12/6/95 & WHT & ISIS\\
            & 1  & 6/11/97 & INTa & IDS\\
            & 1  & 22/11/97 & WHT & ISIS\\
            & 2  & 15/4/03 & INTb & IDS\\
\hline
\dma        & 15 & 29/2-3/3/96 & AAT & RGO\\
(B = 13.05) & 5  & 18-20/3/97 & AAT & RGO\\
\hline
\dmb        & 4$^*$ & 10/6/96 & WHT & UES\\
(B = 15.00) & 6  & 21-25/6/95 & INTa & IDS\\
            & 6  & 11/7/98 & WHT & ISIS\\
            & 8  & 14/7/98 & WHT & ISIS\\
            & 7  & 6-7/10/98 & WHT & ISIS\\
\end{tabular}
\end{table}

AAT: denotes data taken with the Royal Greenwich Observatory (RGO)
spectrograph at the 4\,m Anglo-Australian Telescope (AAT). The setup
consisted of the 82\,cm camera with the R1200R grating centred in
\ha. The CCD used was an MIT-LL (3kx1k) in fast readout mode. This
combination gives a dispersion of 0.23 \app.

INTa: denotes data taken with the Intermediate Dispersion Spectrograph
(IDS) at the 2.5m Isaac Newton Telescope (INT) on the island of La
Palma. For these data, the setup consisted of the 500\,mm camera with
the R1200R grating centred in \ha\ and the Tek (1kx1k) chip. This
combination results in a dispersion of 0.39 \app.

INTb: denotes data taken also with the IDS at the INT but with a setup
that consisted of the 235\,mm camera with the R1200B grating and the
EEV10 (2kx4k) CCD. The wavelength range covered in this case included
\hg\ and \hb. This combination results in a dispersion of 0.48 \app. 

WHT: denotes spectra taken with the 4.2m William Herschel Telescope
(WHT) on La Palma. Most of the spectra were obtained using the double
arm spectrograph ISIS. Only the red spectra were used in our study.
Except for the spectra taken in January 2003, the setup for the red
arm consisted of the 500\,mm camera with the R1200R grating and a Tek
CCD (1kx1k) giving a dispersion of 0.40 \app. For the spectra of \ddb\
taken in January 2003, a MARCONI CCD (2kx4.7k) was used giving a
dispersion of 0.23 \app. In the case of the 4 high resolution spectra
taken of \dmb\ (marked with a star in Table~\ref{obs}), the Utrecht
Echelle spectrograph (UES) was used with the 35 cross disperser and a
Tek (1kx1k) CCD giving a typical dispersion of 0.07 \app.

VLT: denotes data taken with the UV-Visual Echelle Spectrograph (UVES)
at the UT2 (Kueyen) VLT 8.2\,m telescope located at Paranal
Observatory. The setup used consisted of Dichroic~1 (central
wavelengths 3900\,\AA\ and 5640\,\AA) with a EEV CCD (2kx4k) for the
blue arm and two CCDs, a EEV (2kx4k) and a MIT-LL (2kx4k) for the red
arm. This setup allows us to achieve almost complete spectral coverage
from 3200\,\AA\ to 6650\,\AA\ with only two $\sim$80\,\AA\ wide gaps
at 4580\,\AA\ and 5640\,\AA. A slit width of 2.1'' was used to
minimise slit losses and the CCDs were binned $2\times 2$ to reduce
readout noise. This setup results in a spectral resolution of
0.36\,\AA\ (or better if the seeing disk is smaller than the slit
width) at \ha. Exposure times were 5 min.

3.5\,m: denotes data taken with the double beam TWIN spectrograph at
the 3.5\,m telescope in the Calar Alto Observatory.  Only the red
spectra were used in this paper. The setup for the red arm consisted
of the 230\,mm camera with the T06 grating (1200 grooves mm$^{-1}$)
and a SITe-CCD (2kx0.8k) giving a dispersion of 0.55 \app.

The slit width was set to values between 0.8 and 1 arcsec depending on
the seeing. We made sure in every case that the star filled the slit
to avoid systematic errors in the radial velocities caused by the star
wandering in the slit during an exposure.

For the AAT, INT, and WHT runs we obtained CuAr plus CuNe frames to
calibrate the spectra in wavelength. In the case of the 3.5\,m and VLT
observations the wavelength calibration arc used was a ThAr.  All
target spectra were bracketed by arc spectra taken within one hour and
the wavelength scale interpolated to the time of mid-exposure.  We
subtracted from each image a constant bias level determined from the
mean value in its over-scan region.  Tungsten flatfield frames were
obtained each night to correct for the pixel to pixel response
variations of the chip. Sky flatfields were also obtained to correct
for the pixel to pixel variations of the chip along the slit. After
debiasing and flatfielding the frames, spectral extraction proceeded
according to the optimal algorithm of Marsh \shortcite{m89}. The arcs
were extracted using the profile associated with their corresponding
target to avoid systematic errors caused by the spectra being tilted.
Uncertainties on every point were propagated through every stage of
the data reduction. We did not attempt to correct for light losses in
the slit. For the VLT spectra a special procedure was applied to
correct for a quasi-periodic ripple pattern appearing in many of the
uncorrected merged spectra \cite{nkl05}. The resulting VLT spectra
were then divided by a smoothed spectrum of a DC white dwarf, which by
definition shows no spectral features at all and therefore provides an
excellent means of correcting for the instrumental response.

\section{Results}

\subsection{Average spectra}

\begin{figure*}
\begin{picture}(100,0)(10,20)
\put(0,0){\includegraphics{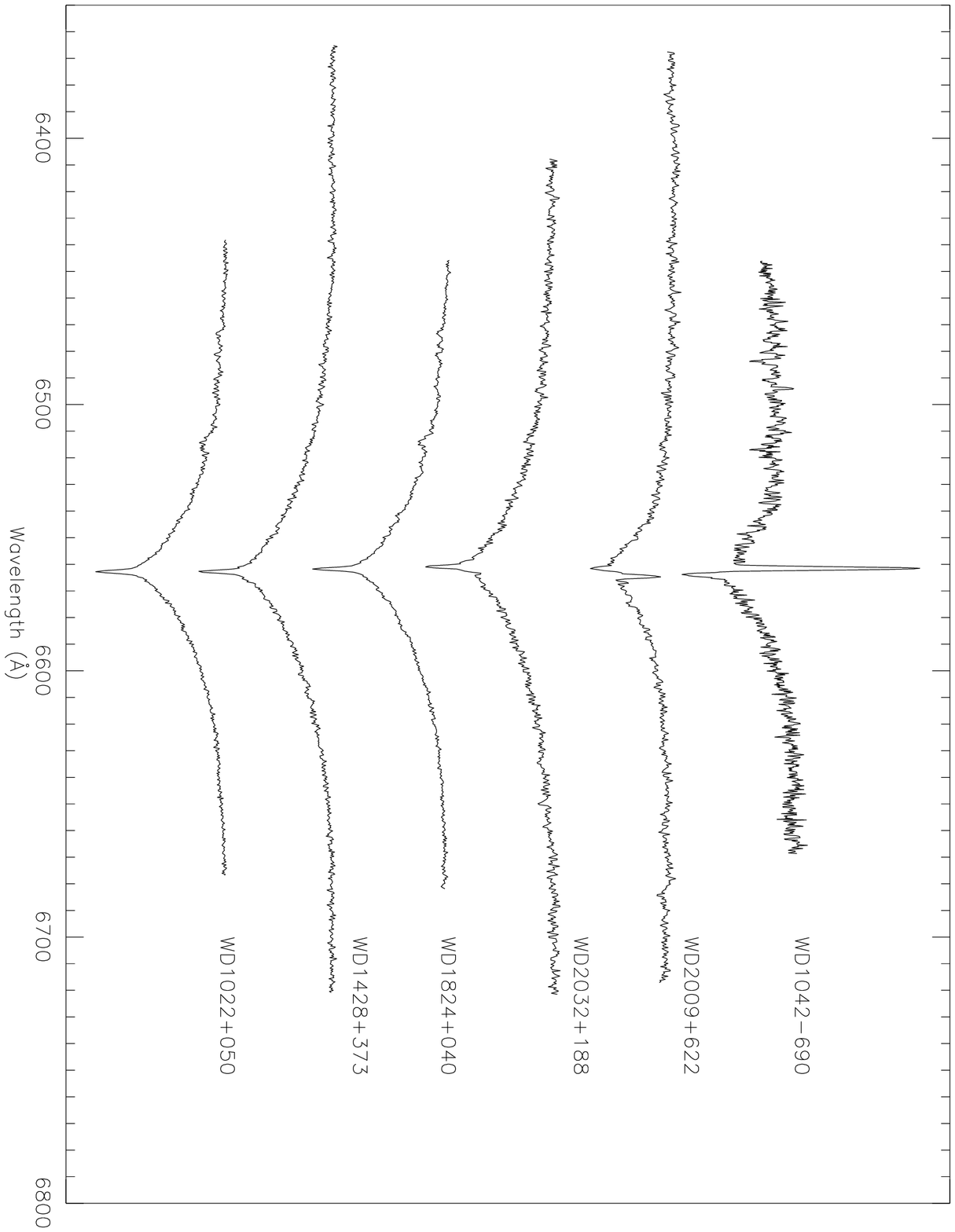}}
\noindent
\end{picture}
\vspace{120mm}
\caption{Average spectra for the six systems studied in this paper.}
\label{res:av}
\end{figure*}

Fig.~\ref{res:av} presents the average red spectra for the six systems
discussed in this paper. The spectra of four of the systems, \dda,
\ddb, \ddc\ and \ddd, are very similar showing only very broad
absorption at \ha. \dma\ and \dmb, on the other hand, show a double
line structure in their \ha\ line profile composed of broad absorption
coming from the white dwarf and narrow emission from the heated
surface of the M dwarf companion.  As this narrow emission moves
within the absorption profile with the orbital period, an average
spectrum would show the emission broadened and for that reason for
these two systems we present the spectrum at a particular orbital
phase instead of an average of the spectra during an orbit. The narrow
emission in \dma\ is, on average, significantly stronger than for
\dmb. For \dmb, the strength of the narrow emission core changes with
orbital phase and only at orbital phase 0.5 (when we are looking
directly into the heated face of the M dwarf) reaches similar strength
to that in \dma. See Section~\ref{mdcomp} for details.

\subsection{Four double degenerate systems}
\label{res:rv}

\begin{table*}
  \caption{Radial velocities for the 5 systems. In the case of the 2
    white dwarf-M dwarf systems the values given are measured from the
    M dwarf.}
  \label{results:rv:tab2}
  \begin{center}
    \begin{tabular}{rrrrrr}
HJD $-$ 2440000 & RV (km s$^{-1}$)& HJD $-$ 2440000 & RV (km s$^{-1}$)& HJD $-$
2440000 & RV (km s$^{-1}$)\\
\hline
\multicolumn{2}{c}{WD1022+050} &
\multicolumn{2}{c}{WD1824+040} &
\multicolumn{2}{c}{WD2032+188} \\
10143.1257 & $-$32.23$\pm$2.55 & 9893.4920 & 96.12$\pm$ 3.76 & 9162.5641 & 42.79$\pm$5.58\\
10143.1447 & $-$34.08$\pm$3.03 & 9893.4999 & 95.64$\pm$ 4.97 & 9162.5752 & 52.97$\pm$5.51\\
10144.0893 & 20.85$\pm$2.67 & 10143.2756 & 58.98$\pm$ 2.83 & 9162.6371 & 54.49$\pm$4.82\\
10144.1012 & 16.48$\pm$2.46 & 10143.2818 & 56.00$\pm$ 3.20 & 9162.6485 & 53.30$\pm$5.03\\
10145.0577 & 89.35$\pm$1.96 & 10143.2880 & 52.64$\pm$ 2.81 & 9214.5046 & 97.72$\pm$4.18\\
10145.0697 & 81.96$\pm$2.01 & 10143.2949 & 51.37$\pm$ 3.05 & 9214.5220 & 99.70$\pm$4.15\\
10146.0863 & 111.94$\pm$2.21 & 10144.2766 & 100.47$\pm$ 1.85 & 9214.5414 & 102.52$\pm$4.47\\
10146.0983 & 116.99$\pm$2.04 & 10144.2851 & 100.91$\pm$1.96 & 9878.6788 & $-$7.75$\pm$4.36\\
10146.1719 & 99.17$\pm$1.99 & 10144.2936 & 102.73$\pm$2.10 & 9878.6848 & $-$5.62$\pm$4.46\\
10146.1839 & 97.42$\pm$2.22 & 10145.2760 & 101.46$\pm$2.65 & 9879.7064 & 62.70$\pm$8.39\\
10527.1487 & 5.75$\pm$3.16 & 10145.2892 & 99.27$\pm$3.30 & 9879.7136 & 61.99$\pm$7.84\\
10527.1584 & 8.34$\pm$3.63 & 10146.2791 & 58.95$\pm$2.54 & 9879.7207 & 72.30$\pm$8.73 \\
10527.1670 & 5.32$\pm$3.83 & 10146.2862 & 57.30$\pm$2.88 & 9880.7100 & 98.97$\pm$7.60\\
10852.5812 & $-$31.10$\pm$4.68 & 10146.2934 & 57.87$\pm$3.59 & 9880.7171 & 95.36$\pm$7.19\\
10852.5911 & $-$41.51$\pm$3.72 & 10146.2986 & 54.33$\pm$8.30 & 9880.7270 & 98.67$\pm$7.50\\
10854.4604 & 56.58$\pm$4.83 & 10526.2345 & 89.86$\pm$1.74 & 10759.3259 & 57.43$\pm$8.35\\
10854.4688 & 66.99$\pm$4.17 & 10526.2465 & 92.02$\pm$1.68 & 10775.3924 & 97.21$\pm$3.23\\
10855.4728 & 109.75$\pm$3.07 & 10526.2585 & 96.23 $\pm$1.81 & 12745.6895 & $-$39.02$\pm$7.58\\
10855.4853 & 109.26$\pm$2.80 & 10526.2715 & 95.24$\pm$1.64 & 12745.7036 & $-$22.83$\pm$6.93\\
10969.8494 & 102.51$\pm$3.08 & 10526.2835 & 94.06$\pm$1.60 & \multicolumn{2}{c}{WD1042$-$690}\\
10969.8567 & 107.65$\pm$3.09 & 10526.2955 & 92.74$\pm$1.36 & 10143.1591 & 76.49$\pm$0.39\\
12744.4589 & $-$30.08$\pm$4.44 &  10526.3078 & 97.26$\pm$2.34 & 10143.1687 & 76.13$\pm$0.36\\
12744.4731 & $-$40.03$\pm$4.63 &  10527.2658 & 109.40$\pm$4.04 & 10143.9950& $-$56.70$\pm$0.44\\
12746.5091 &    26.01$\pm$3.96  & 10527.2777 & 113.65$\pm$3.27 & 10144.0000& $-$57.84$\pm$0.44\\
12746.5232 &    23.86$\pm$4.21  & 10527.2898 & 109.12$\pm$2.87 & 10144.1363 & 57.57$\pm$0.54\\
\multicolumn{2}{c}{WD1428+373} & 10527.2995 &109.09$\pm$4.06 & 10144.1401 & 61.33$\pm$0.56\\
10495.7128 & $-$40.29$\pm$7.71  & 10528.2664 & 63.33$\pm$3.97 & 10144.2244 & 54.00$\pm$0.48\\
10495.7230 & $-$29.57$\pm$ 8.36 & 10528.2809 & 69.52$\pm$3.61 & 10144.2586 & 9.87$\pm$0.54\\
10853.7482 & $-$7.31$\pm$ 3.91  & 10529.2823 & 10.56$\pm$42.73& 10144.2625 & 5.78$\pm$0.57\\
10853.7658 & $-$10.22$\pm$ 3.99 & 10529.2931 & 9.85$\pm$2.60 & 10144.2663 & $-$1.43$\pm$0.67\\
10854.6999 &    40.78$\pm$ 3.68 & 10529.3025 & 5.22$\pm$3.03 & 10144.2702 & $-$3.85$\pm$0.57\\
10854.7709 &    34.01$\pm$ 3.39 & 10529.3095 & 12.36$\pm$5.84 & 10145.0378 & $-$58.52$\pm$0.55\\
11240.5750 & $-$79.23$\pm$ 6.07 & 10622.5164 & 50.08$\pm$3.68 & 10145.0416 & $-$56.30$\pm$0.47\\
11240.6267 & $-$65.38$\pm$ 3.68 & 10623.7062 & $-$6.21$\pm$4.38 & 10145.9757 & $-$29.74$\pm$0.58\\
11240.7237 & $-$11.07$\pm$ 3.80 & 11681.7165 & 35.50$\pm$1.12 & 10145.9808 & $-$34.51$\pm$0.57\\
11241.5861 & $-$81.31$\pm$ 3.88 & 11682.8933 & $-$15.40$\pm$0.79 & 10526.1651 & 20.20$\pm$0.47\\
11241.6592 & $-$89.18$\pm$ 3.36 & 12033.8514 & $-$13.50$\pm$1.12 & 10527.1754 & 21.97$\pm$0.45\\
11241.7579 & $-$54.58$\pm$ 4.47 & 12078.7265 & 6.4$\pm$0.63 & 10528.1260 & 77.15$\pm$0.75\\
11242.5826 & $-$57.44$\pm$ 3.76 & 12116.5747 & 21.30$\pm$0.52 & 10528.1406 & 69.36$\pm$0.92\\
11242.7032 & $-$88.63$\pm$ 3.89 & 12117.5939 & 81.90$\pm$0.56 & 10528.1558 & 59.51$\pm$1.04\\
11242.7749 & $-$92.46$\pm$ 4.03 & 12139.5163 & 13.20$\pm$2.44 & \multicolumn{2}{c}{WD2009+622}\\
11749.3945 & $-$95.69$\pm$ 1.98 & 12209.3141 & $-$12.86$\pm$1.41 & 9889.6260 & $-$25.63$\pm$3.04\\
11749.4086 & $-$88.57$\pm$ 2.23 & 12209.3241 & $-$15.21$\pm$1.51 & 9891.6758 & 50.04$\pm$3.81\\
11749.4247 & $-$95.29$\pm$ 2.09 & 12210.3159 & 7.46$\pm$1.67 & 9892.6414 & $-$93.00$\pm$3.41\\
11749.4388 & $-$91.40$\pm$ 2.05 & 12210.3322 & 14.18$\pm$1.04 & 9892.6633 & $-$123.83$\pm$3.55\\
11749.4580 & $-$84.93$\pm$ 2.05 & 12210.3579 & 12.56$\pm$1.59 & 9893.6384 & $-$208.76$\pm$4.27\\
11749.4721 & $-$79.71$\pm$ 2.10 & 12211.3070 & 60.44$\pm$1.68 & 9893.6603 & $-$188.42$\pm$3.83\\
11749.4885 & $-$73.10$\pm$ 2.28 & 12211.3198 & 65.92$\pm$1.34 & 10244.6034 & 32.39$\pm$1.61\\
11749.5026 & $-$70.89$\pm$ 2.25 & 12211.3344 & 69.37$\pm$1.27 & 10244.6270 & 12.99$\pm$1.63\\
12663.7233 & 40.86$\pm$4.25 & 12212.3669 & 110.82$\pm$1.33 & 10244.6798 & $-$38.75$\pm$1.68\\
12663.7801 & 42.97$\pm$4.43 & 12213.3830 & 80.96$\pm$2.12 & 10244.7081 & $-$77.49$\pm$1.73\\
\multicolumn{2}{c}{WD1824+040} & 12328.7068 &$-$18.31$\pm$1.22 & 11005.7017 & 27.46$\pm$2.33\\
9887.5980 & 91.57$\pm$ 12.09 & 12329.6908 & 22.18$\pm$2.56 & 11005.7077 & 35.25$\pm$2.31\\
9887.6209 & 99.20$\pm$ 14.57 & 12330.7425 & 86.05$\pm$1.75 & 11005.7136 & 38.70$\pm$2.18\\
9887.6529 & 106.03$\pm$ 4.25 & 12331.7437 & 102.40$\pm$3.32 & 11005.7213 & 33.43$\pm$2.22\\
9888.5371 & 95.16$\pm$ 2.44 & 12332.7505 & 85.59$\pm$3.65 & 11005.7272 & 44.31$\pm$2.30\\
9888.5455 & 94.28$\pm$ 3.47 & \multicolumn{2}{c}{WD2032+188} & 11005.7324 & 33.92$\pm$3.87\\
9889.5252 & 45.32$\pm$ 2.24 & 9150.6424 &$-$40.70$\pm$6.83 & 11008.6761 & 34.63$\pm$2.64\\
9889.5332 & 41.20$\pm$ 3.19 & 9150.6533 &$-$18.85$\pm$6.08 & 11008.6820 & 41.52$\pm$2.57\\
9891.5276 & $-$2.10$\pm$ 2.48 & 9150.6677 &$-$21.41$\pm$6.37 & 11008.6879 & 41.85$\pm$2.57\\
9891.5356 & $-$0.29$\pm$ 3.63 & 9150.6788 &$-$20.83$\pm$7.99 & 11008.6939 & 40.15$\pm$2.60\\
9892.5007 & 41.05$\pm$ 4.06 & 9153.6509 & 95.30$\pm$11.16 & 11008.7012 & 44.00$\pm$2.78\\
9892.5087 & 37.97$\pm$ 3.87 & 9153.6702 & 100.11$\pm$9.50 & 11008.7072& 51.42$\pm$2.76\\

   \end{tabular}
  \end{center}
\end{table*}
\setcounter{table}{1}
\begin{table*}
  \caption{Continued.}
  \label{results:rv:tab2b}
  \begin{center}
    \begin{tabular}{rrrrrr}
HJD - 2440000 & RV (km s$^{-1}$)& HJD - 2440000 & RV (km s$^{-1}$)& HJD -
2440000 & RV (km s$^{-1}$)\\
\hline
\multicolumn{2}{c}{WD2009+622} & & & & \\
11008.7131& 42.67$\pm$2.73 & & & \\
11008.7190& 50.81$\pm$2.59 & & & \\
11093.4111& $-$76.67$\pm$3.14 & & & \\
11093.4748& $-$146.25$\pm$5.21 & & & \\
11093.5206& $-$194.63$\pm$3.10 & & & \\
11093.5673& $-$224.30$\pm$3.29 & & & \\
11094.3644& $-$232.52$\pm$2.78 & & & \\
11094.4819& $-$164.87$\pm$2.82 & & & \\
11094.5644& $-$50.85$\pm$3.58 & & & \\
   \end{tabular}
  \end{center}
\end{table*}

\begin{figure*}
\begin{picture}(100,0)(10,20)
\put(0,0){\includegraphics{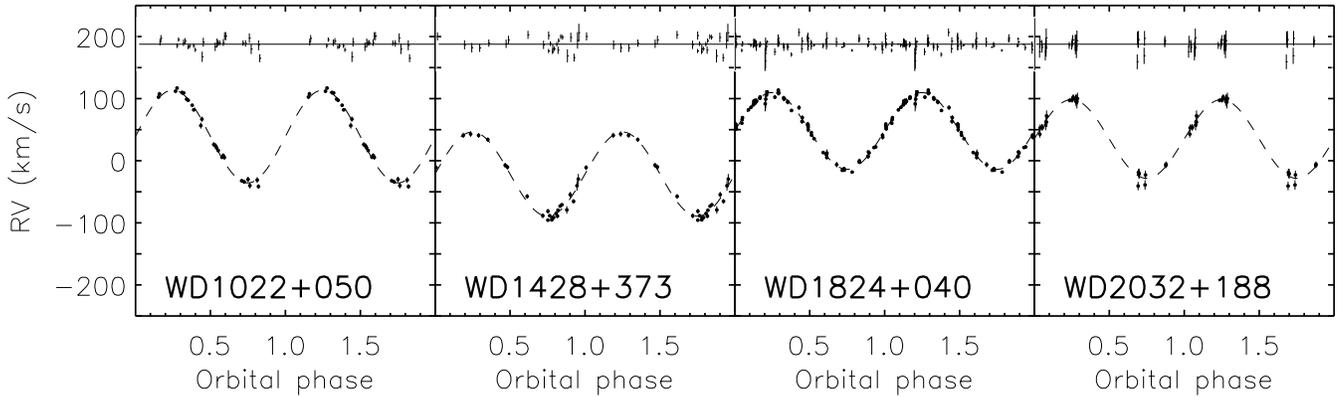}}
\noindent
\end{picture}
\vspace{65mm}
\caption{Radial velocity curves for the four double degenerate
  systems. The data have been folded on the orbital period in each
  case. See Table~\ref{res:rv:tab1} for the list of periods, radial
  velocity semiamplitudes and systemic velocities. Included in each
  panel is a plot of the residuals to the fit. The vertical scale on
  which the residuals have been plotted is twice the scale on which
  the radial velocities are plotted.}
\label{res:rv:rvWD}
\end{figure*}

\begin{figure*}
\begin{picture}(100,0)(10,20)
\put(0,0){\includegraphics{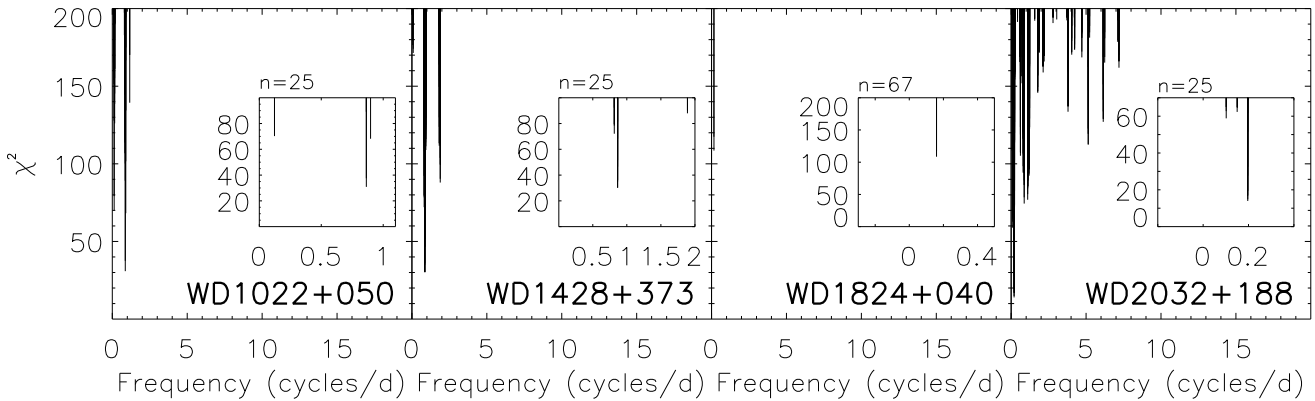}}
\noindent
\end{picture}
\vspace{65mm}
\caption{Each panel presents $\chi^2$ versus cycles/day obtained after
the period search was carried out. The frequency with the smallest
$\chi^2$ corresponds to the orbital frequency of the system. For
clarity we have also included an inset showing a blow up of the region
where the best period is. The number of radial velocity measurements
used for the period search calculations, n, is shown in each panel.}
\label{res:rv:pgramWD}
\end{figure*}

\begin{table*}
\caption{List of the orbital periods measured for the four double
  degenerate systems studied. T$_{0}$, the systemic velocity,
  $\gamma$, the radial velocity semi-amplitude, K, the reduced
  $\chi^2$ achieved for the best alias, the 2nd best alias and the
  $\chi^2$ difference between the 1st and 2nd aliases are also
  presented. When calculating the $\chi^2$ for both aliases we have
  added in quadrature a systematic error that results in a reduced
  $\chi^2 \sim$1 (see text for details). The number of data points
  used to calculate the orbital period is given in the final column
  under N.}
\label{res:rv:tab1}
\begin{center}
\begin{tabular}{lllccllll}
Object & HJD (T$_{0}$) & Period (d)& $\gamma$ (km/s) & K (km/s) &
$\chi^{2}_{reduced}$ &2nd best alias (d)& $\Delta\chi^{2}$ & N \\

& $-$2400000 & & &  & & & \\
\hline

WD1022+050 & 51445.262(5) & 1.157155(5) & 39.05$\pm$1.19
&74.77$\pm$1.16 & 1.25 & 8.1580(4) & 40 & 25\\

WD1428+373 & 51579.64(1) & 1.15674(2) & $-$21.46$\pm$1.62
&67.90$\pm$1.68 & 1.41 & 1.22640(1) & 42 & 25\\

WD1824+040 & 51108.192(9) & 6.26600(5) & 47.95$\pm$0.40
& 61.87$\pm$0.55 & 1.24 & 0.5449639(5) & 3064 & 67\\

WD2032+188 & 50947.07(5) & 5.0846(3) & 35.11$\pm$1.52 
& 63.50$\pm$1.59 & 0.66 & 9.8267(6) & 45 & 25\\

\end{tabular}
\end{center}
\end{table*}

To measure the radial velocities of the four double degenerates:
WD1022+050, \ddb, WD1824+040 and WD2032+188, we used least squares
fitting of a model line profile. The model line profile is the
summation of three Gaussian profiles with different widths and depths.
For any given star, the widths and depths of the Gaussians are
optimised and then held fixed while their velocity offsets from the
rest wavelengths of the lines in question are fitted separately for
each spectrum; see Maxted, Marsh \& Moran \shortcite{m00c} for further
details of this procedure.

Once the radial velocities for each system were known (see
Table~\ref{results:rv:tab2}) we used a ``floating mean'' periodogram
to determine the periods of our targets (e.g. Cumming, Marcy \& Butler
1999). The method consists in fitting the data with a model composed
of a sinusoid plus a constant of the form:
\[  \gamma + {\rm K}  \sin (2 \pi f (t - t_0)),\]
where $f$ is the frequency and $t$ is the observation time. The key
point is that the systemic velocity is fitted at the same time as $K$
and $t_0$.  This corrects a failing of the well-known Lomb-Scargle
\cite{l76,s82} periodogram which starts by subtracting the mean of the
data and then fits a plain sinusoid; this is not the best approach for
small numbers of points. We obtained the $\chi^2$ of the fit as a
function of $f$ and then identified minima in this function.

Table \ref{res:rv:tab1} gives a list of the orbital parameters derived
for each DD binary star. The orbital period of the second best alias
is also given, along with the difference in $\chi^2$ between the two
best periods found. The large difference in $\chi^2$ indicates that
the second best aliases are not plausible. The resulting radial
velocity curves (folded in the orbital period) are presented in
Fig.~\ref{res:rv:rvWD} and the corresponding periodograms ($\chi^2$
versus orbital frequency) in Fig.~\ref{res:rv:pgramWD}. Each panel in
the periodogram includes a blow up of the region in frequency where
the minimum $\chi^2$ is found.

In each case, we compute the level of systematic uncertainty that when
added in quadrature to our raw error estimates gives a reduced $\chi^2
\sim 1$. By doing this we are considering the un-accounted sources of
error such as true variability of the star or slit-filling errors that
cause the poor fits of a few stars. Such errors are unlikely to be
correlated with either the orbit or with the statistical errors we
estimate, and therefore we add a fixed quantity in quadrature with our
statistical errors as opposed to applying a simple multiplicative
scaling to them. In all cases we use a minimum value of $2\,{\rm
km}\,{\rm s}^{-1}$ corresponding to 1/10$^{\rm th}$ of a pixel which
we believe to be a fair estimate of the true limits of our data. The
last column of Table~\ref{res:prob} gives the value of systematic
uncertainty that we have added in quadrature in each case.

We then calculate the probability of the true orbital periods being
further than 1 and 10 per cent from the values we obtained - see
Morales-Rueda et al. \shortcite{lmr03} and Marsh, Dhillon \& Duck
\shortcite{mdd95} for an explanation of the method used to calculate
these probabilities - and present them in Table~\ref{res:prob}.

In all cases, the probabilities of the quoted periods being wrong are
very low and we are certain that the values given in
Table~\ref{res:rv:tab1} correspond to the true orbital solution.

In the cases where the probability of the orbital period being further
than 1 and 10 per cent from our favoured value is the same, the
significant probability lies within a very small range around the best
period, with all the significant competition (i.e.  next best alias)
placed outside the 10 per cent region around the best alias.

\begin{table}
\caption{List of probabilities that the true orbital period of a
  system lies further than 1 and 10 per cent from our favoured value
  given in Table~\ref{res:rv:tab1}. Numbers quoted are the logarithms
  in base 10 of the probabilities. Column number 4 gives the value of
  the systematic uncertainty that has been added in quadrature to the
  raw error to give a reduced $\chi^2 \sim$ 1.}
\label{res:prob}
\begin{center}
\begin{tabular}{lccc}
Object & 1\%\ & 10\%\ & systematic error\\
 & & & (km\,s$^{-1}$) \\
\hline
\dda\  & $-$7.66 & $-$7.69 & 3\\
\ddb\  & $-$9.59 & $-$13.08 & 4\\
\ddc\  & $-$1000 & $-$1000 & 2\\
\ddd\  & $-$9.89 & $-$9.90 & 2\\
\end{tabular}
\end{center}
\end{table}

\subsubsection{The unseen component of the binary}
\label{res:unseen}

By knowing the radial velocity semiamplitude of one of the components
of the binary (the observable component), $K$, and the orbital period
of the system, we can then calculate the mass function of the unseen
component by using:
\begin{equation}
\label{eq1}
f_m = \frac{M_2^3 \sin^3 i}{(M_1 +M_2)^2} =
\frac{P K_{1}^3}{2 \pi G},
\end{equation}
where the subscripts ``$1$'' and ``$2$'' refer to the brighter and the
dimmer components respectively. The mass function is the lower limit
of the mass of the unseen component.  Table~\ref{res:mass} gives the
mass functions of the unseen components for the four DDs studied. In
two cases (\ddc\ and \ddd) the companion's mass function is greater
than 0.1\msun\ which corresponds to the mass of a late M dwarf if it
is a main sequence star. The masses of the brighter components of the
systems have been measured by fitting their hydrogen line profiles to
stellar atmosphere models using the tracks by Althaus \& Benvenuto
\shortcite{ab97} and can be substituted, together with the assumption
of the orbital inclination of the system being 90$^\circ$, in the mass
function equation to give a larger lower limit for the masses of the
unseen components. These revised lower limits (also presented in
Table~\ref{res:mass}) are all greater than 0.1\msun\ which indicates
that the unseen companions cannot be main sequence stars because if
they were we should be able to detect them \cite{mdd95}. The unseen
companions must therefore be also compact objects, probably white
dwarfs.

\begin{table}
\caption{The mass functions, $f_m$, of the unseen components together
  with the larger lower limits obtained by assuming $i$ = 90$^\circ$
  and by substituting in the mass function equation our determination
  of the mass of the brighter component, $M_1$.}
\label{res:mass}
\begin{center}
\begin{tabular}{llll}
Object & $f_m$($\msun$) & $M_1$($\msun$)  & $M_2$($\msun$) \\
       &                &                 &lower limit      \\
\hline

WD1022+050 & 0.050 & 0.389 & 0.283 \\
WD1428+373 & 0.038 & 0.348 & 0.233 \\
WD1824+040 & 0.154 & 0.428 & 0.515 \\
WD2032+188 & 0.135 & 0.406 & 0.469 \\

\end{tabular}
\end{center}
\end{table}

\begin{figure*}
\begin{picture}(100,0)(10,20)
\put(0,0){\includegraphics{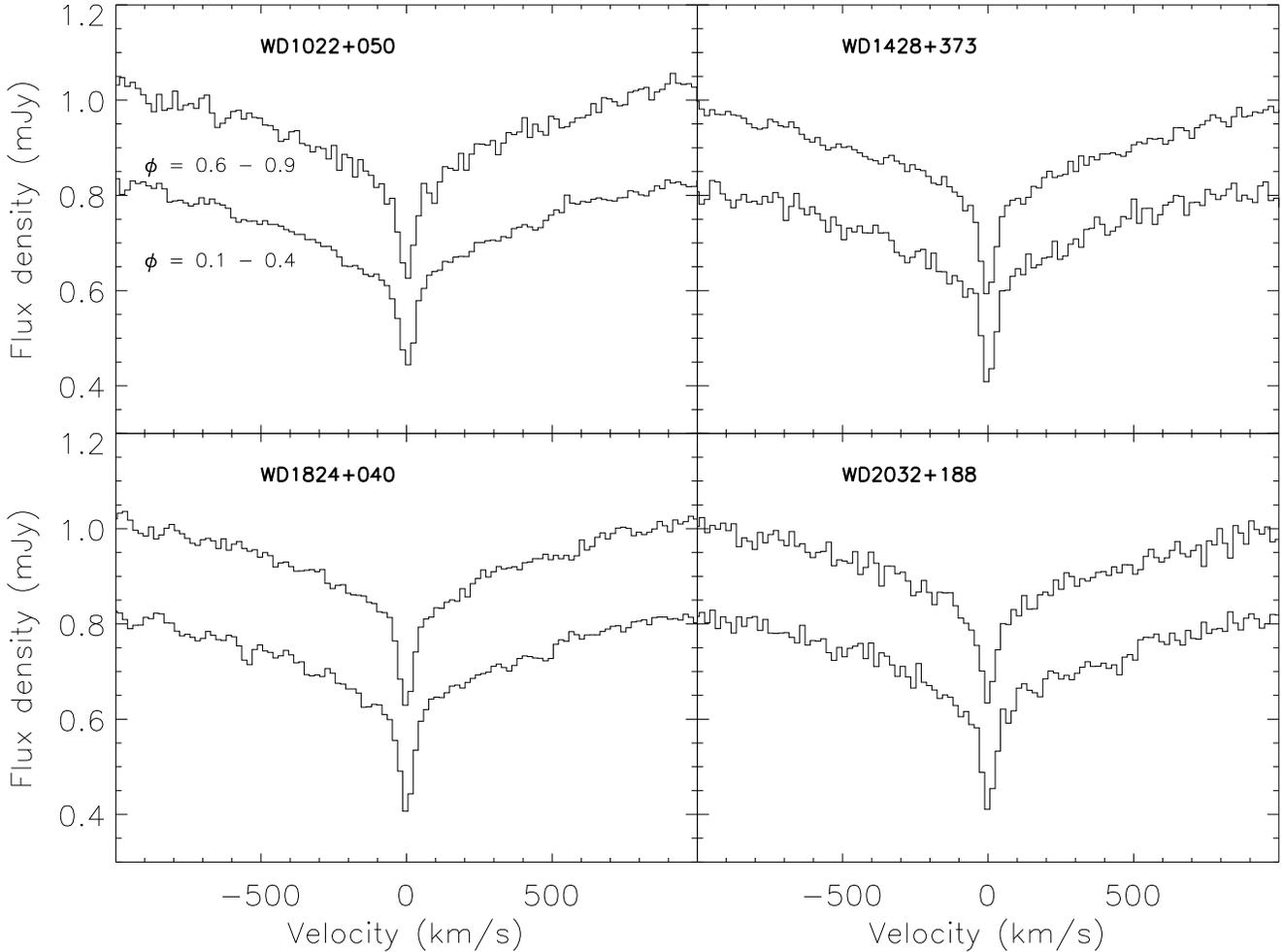}}
\noindent
\end{picture}
\vspace{130mm}
\caption{The spectra averaged around the quadrature phases for the
  four systems studied. In each case, the lower spectrum corresponds
  to quadrature phase 0.25 and the top one to phase 0.75. There is no
  clear asymmetry in the line profile at phase 0.25 that gets mirrored
  in phase 0.75 for any of the systems. This indicates that we cannot
  detect the faint companion of the systems.}
\label{res:comp}
\end{figure*}

We searched for the signature of the faint companions by shifting out
the fitted radial velocity for each binary and then looking for
differences in the line profiles at the quadrature phases
\cite{mdd95}, i.e. 0.25 and 0.75. The spectra at quadrature phases
were obtained by averaging the spectra contained in two separate phase
ranges, i.e. the spectra in the range from 0.1 to 0.4 were averaged to
obtain the phase 0.25 spectrum, and the spectra in the range from 0.6
to 0.9 to obtain the phase 0.75 spectrum. The results are plotted in
Fig.~\ref{res:comp}. Any contribution from the companion white dwarf
should be seen as an asymmetry in the line profile at phase 0.25 that
is mirrored at phase 0.75 with respect to the rest wavelength
\cite{mdd95}. None of the four systems show a clear asymmetry of this
type in the line profiles.

A second test that can be carried out to look for the faint companion
consists in shifting out the fitted radial velocity off the spectra
and creating a mean spectrum by combining all the shifted spectra,
subtracting this mean spectrum from the individual shifted ones and
plotting the resulting spectra in a stack or trail. This aids the eye
to identify any leftover absorption moving with the binary orbit. A
Doppler map \cite{mh88} can also be computed from these stack of
spectra. Any orbital motion leftover in the spectra would appear in
the maps as a small absorption region located in the $V_x = 0$ axis of
the velocity map. We do not find any indication of the presence of the
unseen component in either trails or Doppler maps in any of the four
systems studied.

\subsubsection{How faint is the unseen companion?}

We explore the question of how small the contribution of the faint
component of the system has to be so as not to be detected using the
methods discussed in the previous section.

To answer this question we create synthetic spectra that include the
absorption corresponding to the brighter component of the system (in
the form of three Gaussians), scaled to the measured value for each
system, plus some extra absorption moving opposite to it (also
represented by three Gaussians) and determine for what percentage of
brightness, relative to the bright component, we should be able to
detect the faint component by looking at the spectra around the
quadrature phases \cite{mdd95}. This method assumes that the companion
stars have a spectrum similar to that of the brighter
component. Although the fainter white dwarf is cooler and its \ha\
line will be less deep in its spectrum, this seems like a reasonable
assumption as the companions are also white dwarfs, unless of course
they are not DA white dwarfs.

In the case of \dda\ we find that the brightness of the companion must
be less than 10 per cent the brightness of the bright component for us
not to detect it. The values we find for \ddb, \ddc, and \ddd\ are 9
per cent, 23 per cent and 17 per cent respectively. These values
translate into magnitudes for the companions, in the 6400--6700\AA\
region, that are respectively 2.5, 2.6, 1.6 and 1.9 fainter than that
of the bright components. Using the calculated absolute V magnitudes
for the bright components (10.68, 10.35, 10.83 and 10.24, see the
Discussion Section) and the broadband colour indices for pure hydrogen
and $\log g$ = 8 stellar atmosphere models computed by Bergeron,
Wesemael \&\ Beauchamp \shortcite{bwb95}, we obtain upper limits for
the absolute R magnitudes of the faint components of 13.3, 13.0, 12.5
and 12.2 respectively.

\subsection{Two white dwarf/M dwarf binaries}

\begin{figure*}
\begin{picture}(100,0)(10,20)
\put(0,0){\includegraphics{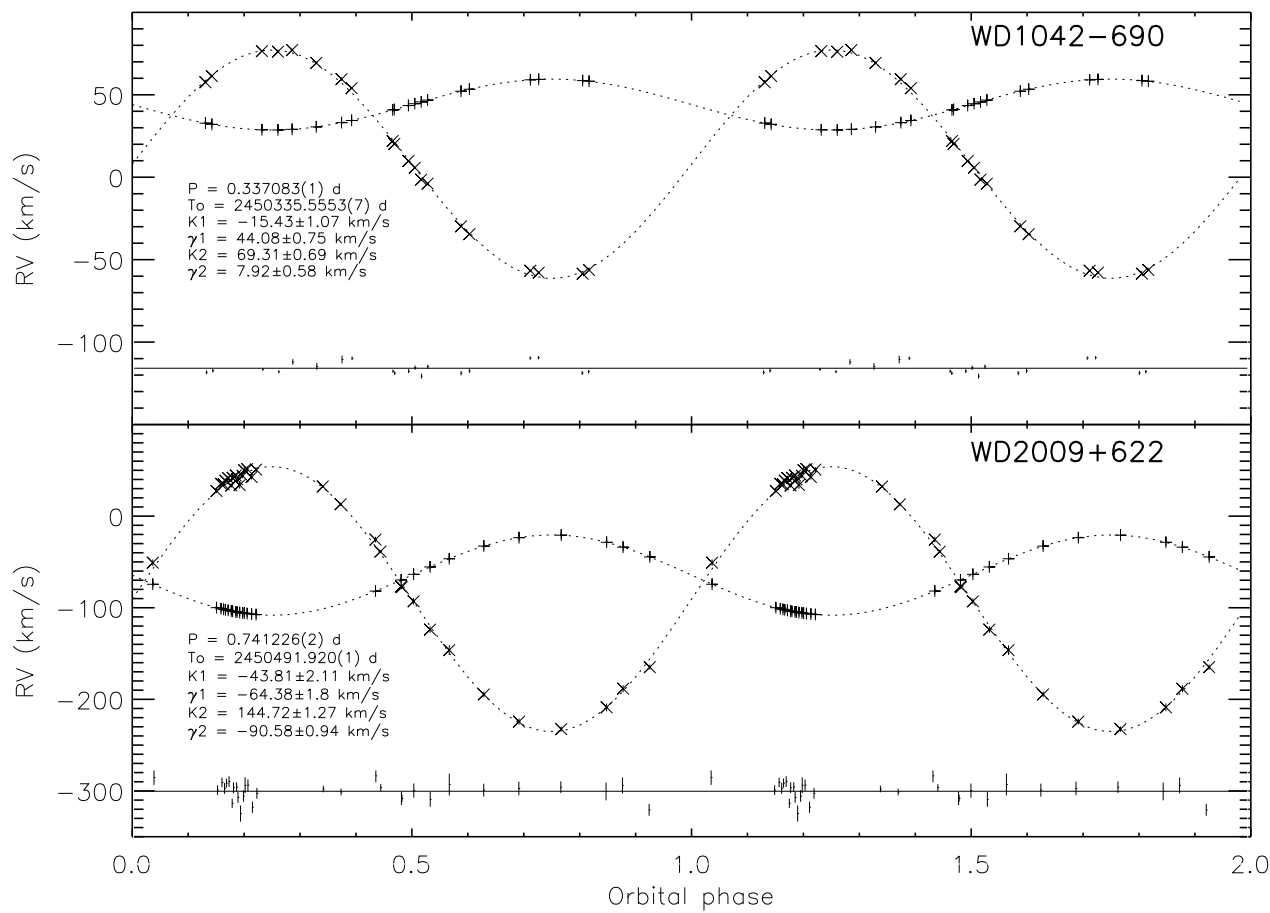}}
\noindent
\end{picture}
\vspace{123mm}
\caption{Orbital solution for \dma\ and \dmb. Included in each panel
  is a plot of the residuals to the fit to the emission line
  component. The vertical scale on which the residuals have been
  plotted is larger than the scale on which the radial velocities are
  plotted.}
\label{res:dadm}
\end{figure*}

In the case of \dma\ (aka BPM 6502) and \dmb\ the spectra are composed
of absorption lines that have their origin in the white dwarf plus
emission cores that have their origin in the M dwarf. This extra
emission component makes the measuring of the radial velocities more
complicated, as the absorption coming from the white dwarf has its
core filled by the M dwarf emission. A way to measure simultaneously
the radial velocities of both components is to use least squares
fitting of a model line profile as in the previous case but this time
using a model line profile that is the sum of four Gaussian profiles.
Three of the Gaussians fit the absorption component and one fits the
emission.  The steps followed to carry out these complex fits
consisted of: 1) fitting only the emission lines with a single
Gaussian function and obtaining the radial velocities associated to
the emission line for each spectrum, 2) calculating the orbital
solution for the emission lines by means of obtaining a periodogram
from the radial velocities measured, 3) using this orbital solution to
fix the orbit of the three Gaussians that will fit the absorption
component of the lines and 4) obtaining the radial velocity
semiamplitude and systemic velocity for the white dwarf by fitting all
the spectra simultaneously with one emission and three absorption
Gaussians.

This method was easily applicable to \dmb\ as the emission coming from
the M dwarf is comparable to the \ha\ absorption core (see
Fig.~\ref{res:av}). In the case of \dma\ the emission component is
very strong compared with the absorption core making the fitting of
the data more difficult and the results obtained less accurate. For
\dma\ we find 7 very close aliases with very similar values of
$\chi^2$. Table~\ref{res:rv:dma} gives a list of the orbital solutions
for the 7 aliases. The solutions for K$_2$, $\gamma_2$, K$_1$ and
$\gamma_1$ are consistent within the errors for all aliases. Previous
studies of \dma\ \cite{k00} result in an orbit solution in which the
orbital period is consistent with our alias number 3 and the values
for K$_2$, $\gamma_2$, K$_1$ and $\gamma_1$ are consistent with those
from our 7 aliases. In Table~\ref{res:rv:tab2} we present the orbital
solutions for \dma\ and \dmb. The results presented for \dma\
correspond to the first alias shown in Table~\ref{res:rv:dma}. In
Fig.~\ref{res:dadm} we plot the radial velocities measured for the
emission and absorption components together with the best fits given
in Table~\ref{res:rv:tab2}. The error bars in the radial velocities
are smaller than the size of the symbols used to plot the data. Notice
that there are 4 extra points in the fit to the radial velocity of the
M dwarf for \dmb. This accounts for the 4 high resolution spectra
taken with UES that only covered the core of the \ha\ line.

\begin{table}
\caption{Orbital solution for the 7 aliases found for
  \dma.}
\label{res:rv:dma}
\begin{center}
\begin{tabular}{lllll}
P (d) & T$_{0}$ & $\gamma_2$ & K$_2$ &
$\chi^{2}_{2red}$ \\
       &$-$2450000 & & & \\
\hline
 0.337083(1) & 335.5553(7) & 7.93$\pm$0.58 & 69.31$\pm$0.69 &
0.65 \\
 0.337380(1) & 335.7239(7) & 7.83$\pm$0.58 & 69.24$\pm$0.69 &
0.74 \\
 0.336786(1) & 335.7239(7) & 8.03$\pm$0.58 & 69.38$\pm$0.69 &
0.75 \\
 0.337678(1) & 335.5550(7) & 7.73$\pm$0.58 & 69.16$\pm$0.69 &
1.01 \\
 0.336490(1) & 335.5557(7) & 8.14$\pm$0.58 & 69.44$\pm$0.70 &
1.03 \\
 0.337977(1) & 335.7238(7) & 7.64$\pm$0.58 & 69.07$\pm$0.68 &
1.47 \\
 0.336194(1) & 335.7240(7) & 8.26$\pm$0.58 & 69.49$\pm$0.70 &
1.48 \\
\end{tabular}
\end{center}
\end{table}

\begin{table}
\caption{List of the orbital periods measured for the two white
  dwarf-M dwarf systems studied. T$_{0}$, the systemic velocity,
  $\gamma$, the radial velocity semi-amplitude, K, for both the white
  dwarf and the M dwarf, and the reduced $\chi^2$ achieved for the
  best alias are given The number of data points used to calculate the
  orbital period is also given.}
\label{res:rv:tab2}
\begin{center}
\begin{tabular}{ccc}
 & \dma\ & \dmb\ \\
\hline
N & 20 & 31/27 \\
P (d) & 0.337083(1) & 0.741226(2) \\
T$_0$ (d) & 2450335.5553(7) & 2450491.920(1)\\
K$_{WD}$ (\kmsec) & $-$15.43$\pm$1.07 & $-$43.81$\pm$2.11\\
$\gamma_{WD}$ (\kmsec) & 44.08$\pm$0.75 & $-$64.38$\pm$1.80\\
K$_{M}$ (\kmsec)& 69.31$\pm$0.69 & 144.72$\pm$1.27\\
$\gamma_{M}$ (\kmsec)& 7.92$\pm$0.58 & $-$90.58$\pm$0.94\\
$\chi^2_{2red}$ & 0.65 & 1.12 \\
q=M$_{M}$/M$_{WD}$ & 0.223$\pm$0.018 & 0.303$\pm$0.017\\
\end{tabular}
\end{center}
\end{table}

\subsubsection{The M dwarf companions}
\label{mdcomp}

\begin{figure*}
\begin{picture}(100,0)(10,20)
\put(0,0){\includegraphics{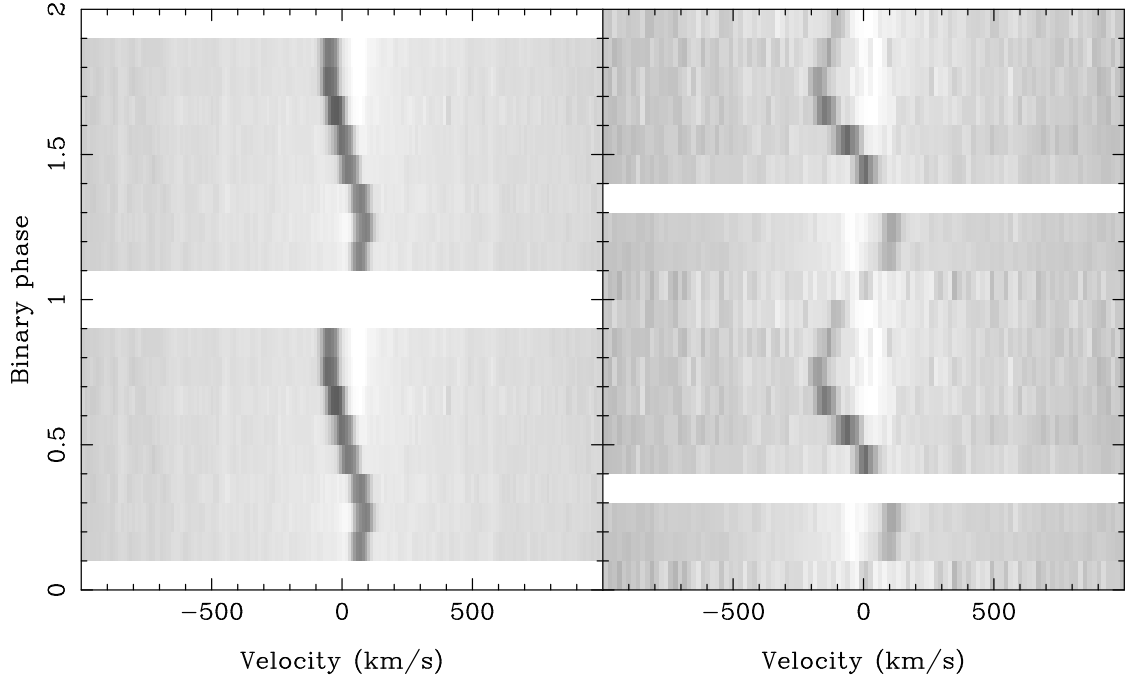}}
\noindent
\end{picture}
\vspace{100mm}
\caption{Trailed phase binned spectra for \dma\ and \dmb. The orbit is
  plotted twice. The variable brightness in the emission line
  component, coming from the M dwarf companion, is clear in \dmb.}
\label{res:wdMdtr}
\end{figure*}

We binned the spectra for both binary systems into 10 phase bins and
plotted them in Fig.~\ref{res:wdMdtr} as a stack of spectra with
orbital phase in the vertical axis. The advantage of presenting the
spectra in this way is that we can explore the line flux variations
during an orbit. In the case of \dma\ the strength of the emission
line does not vary significantly with orbital phase. This is not the
case for \dmb\ where the line flux increases at orbital phases around
0.5. This phase corresponds to the white dwarf and the M dwarf being
aligned with the line of sight, the white dwarf being closer to us. At
this phase we are looking to the heated face of the M dwarf.

\begin{figure}
\begin{picture}(100,0)(-270,209)
\put(0,0){\includegraphics{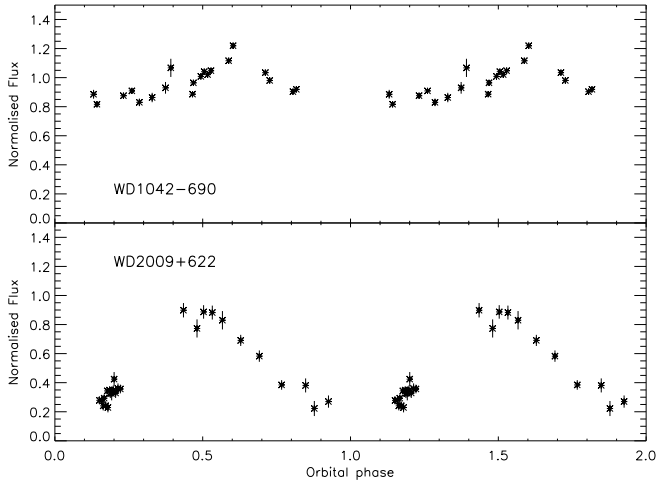}}
\noindent
\end{picture}
\vspace{65mm}
\caption{Flux modulation seen on the emission from the M dwarf
  component of \dma\ (top) and \dmb\ (bottom).}
\label{res:Mdflux}
\end{figure}

\begin{figure}
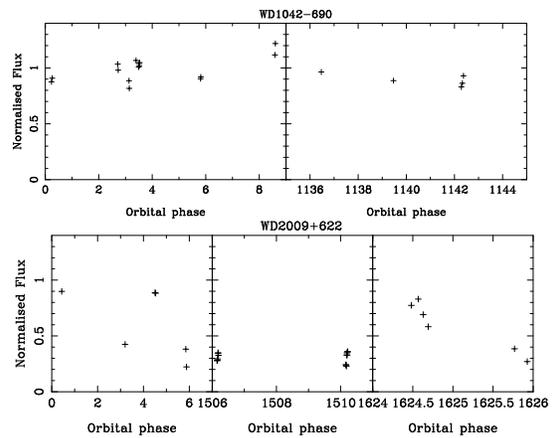

\begin{picture}(100,0)(0,0)
\put(0,0){\includegraphics{1042lc.ps}}
\put(0,0){\includegraphics{2009lc.ps}}
\noindent
\end{picture}
\vspace{60mm}
\caption{Flux modulation seen on the emission from the M dwarf
  component of \dma\ (top) and \dmb\ (bottom). This time the data has
  not been folded in the orbital period.}
\label{res:Mdlc}
\end{figure}

To determine how the line flux varies with phase, we fitted the line
profiles once more with four Gaussians, three for the absorption
component and one for the emission component but this time we allowed
the height of the emission component to vary. The results are
displayed in Fig.~\ref{res:Mdflux}. The variation in flux is very
significant in \dmb\ and as we mentioned earlier it can be explained
with emission coming from the irradiated face of the M dwarf, the side
facing the white dwarf. This has important consequences for the
calculation of the mass of the white dwarf carried out in the
following sections as we have to account for distortions on the M
dwarf to correct its radial velocity semiamplitude. When the companion
is heavily irradiated the emission line used to measure its radial
velocity will give us the value associated with the irradiated face,
not the value for the centre of mass of the companion. Maxted et
al. \shortcite{m98} find that probably as a result of optical depth
effects, Balmer emission lines induced by irradiation are
significantly broadened rendering measurements of radial velocities
from fitting these lines more inaccurate.

In the case of \dma, there is also some flux modulation present at a
fainter level but peaking at phase 0.6 rather than phase 0.5. This is
probably associated to chromospheric activity of the M dwarf rather
than irradiation. To make sure that this variability is not orbital we
have also plotted the flux of the emission component versus unfolded
orbital phase in Fig.~\ref{res:Mdlc}. Wee see that for \dma\ the
variability observed is not phase dependent implying that is
intrinsic to the M dwarf, probably chromospheric. In the case of
\dmb\ the variability observed is larger and it peaks at phase 0.5
confirming that it is due to irradiation of the inner face of the M
dwarf by the white dwarf.

\subsubsection{The masses of the components}
\label{sec:mass}

Once the radial velocity semiamplitudes for both components have been
measured, together with the orbital period we can use Eq.~\ref{eq1} to
obtain the mass function for the white dwarf and the M dwarf in each
case. If we combine Eq.~\ref{eq1} with $q = M_{M}/M_{WD} =
K_{WD}/K_{M}$ we obtain larger lower limits for the masses of both
components:

\begin{eqnarray}
\label{mass}
M_{WD} = \frac{PK_{M}(K_{WD} + K_{M})^2}{2\pi G \sin^3{i}},\nonumber \\
M_{M} = \frac{PK_{WD}(K_{WD} + K_{M})^2}{2\pi G \sin^3{i}}.
\end{eqnarray}

The actual mass of the white dwarf can be determined from the
gravitational redshift and the mass-radius relationship for white
dwarfs \cite{ab97}. First we calculate the difference in systemic
velocities for both system components ($\gamma_{WD} - \gamma_{M}$). We
must then add corrections for (i) the redshift of the M dwarf,
$GM_{M}/R_{M}c$, where the radius of the M dwarf has been calculated
by using the mass-radius relationship given by Caillault \& Patterson
\shortcite{cp90}:
\[\log{R/\rsun} = 0.796 \log{M/\msun} - 0.037,\]
(ii) the difference in transverse Doppler shifts of both components:
\[(K^2_{M} - K^2_{WD})/2 c \sin^2{i},\]
(iii) the potential at the M dwarf produced by the white dwarf:
\[G M_{WD}/ac,\]
(iv) and the potential at the white dwarf due to the M dwarf:
\[G M_{M}/ac,\]
where $a$ is the distance between both stars. The mass of the white
dwarf is then calculated by comparing the resulting gravitational
redshift with models for low mass helium white dwarfs
\cite{ab97,ba98}. The inclination of the system can then be calculated
by using Eq.~\ref{mass}.

In order to calculate the corrections to the white dwarf's
gravitational redshift we had to assume a value for the mass of the M
dwarf. We performed several iterations of these calculations until we
obtained consistent values for all the parameters. This method has
been used previously \cite{md96,m98} to obtain the masses of both
components in pre-CV systems. The results obtained after these
iterations are presented in Table~\ref{res:wdMd:mass}.

We chose as the best estimates for the corrections described above and
the final parameters of the iterations, those that resulted in a mass
for the white dwarf that was closer to that calculated by Benvenuto
\&\ Althaus \shortcite{ba98}. A minimum value for the masses of the
components was found when we assumed a helium core white dwarf, with
metallicity Z = 0.001 and an outer hydrogen envelope of fractional
mass (i.e. mass of envelope/total mass of the star) 10$^{-8}$. We
assumed a T$_{\rm eff}$ = 21380 K and 25870 K for \dma\ and \dmb\
respectively \cite{brb95,bsl92}. A maximum value was found if instead
we assumed that the fractional mass of the outer hydrogen envelope was
4$\times$10$^{-4}$ in the case of \dma\ and 2$\times$10$^{-4}$ in the
case of \dmb. The minimum and maximum values found for the masses are
given in Table ~\ref{res:wdMd:mass}.  

\begin{table}
\caption{Summary of the parameters for \dma\ and \dmb. BA indicates
  Benvenuto \&\ Althaus \protect\shortcite{ba98}. See text for explanations on
  the different values given.}
\label{res:wdMd:mass}
\begin{tabular}{lll}
Parameter & white dwarf & M dwarf \\
\hline
\dma \\
\hline
f$_m$(\msun) & 0.0116(2) & 0.00013(1)\\
M(\msun) from Eq~\ref{mass} & 0.0174(9) & 0.0039(4)\\
M(\msun) from q & 0.75(5)/0.78(5) & 0.1665(5)/0.1735(5)\\
M(\msun) from BA & 0.75(7)/0.78(7) & \\
M(\msun) if CO WD & 0.72 \\
i ($^\circ$) & 16 &  \\
\hline
\dmb \\
\hline
f$_m$(\msun) & 0.233(4) & 0.0065(5)\\
M(\msun) from Eq~\ref{mass} & 0.40(3) & 0.12(2)\\
M(\msun) from q & 0.61(3)/0.64(3) & 0.1845(5)/0.1925(5)\\
M(\msun) from BA & 0.61(3)/0.64(3) &\\
M(\msun) if CO WD & 0.59 \\
i ($^\circ$) &60/59 & \\
\end{tabular}
\end{table}

For \dma, $\gamma_{WD} - \gamma_{M}$ is 36.16$\pm$2.72\kmsec.  After
iterating for different values for the mass of the M dwarf we find
that if M$_{\rm M}$ is 0.166 - 0.167 \msun\ the corrections i, ii, iii
and iv are respectively 0.48, 0.09, 0.24 and $-$0.05 \kmsec, giving a
value for the white dwarf redshift of 36.92 \kmsec. If, on the other
hand, M$_{\rm M}$ is 0.173 - 0.174 \msun\, the corrections are
respectively 0.48, 0.10, 0.25, $-$0.05 \kmsec, giving a value for the
white dwarf redshift of 36.93 \kmsec.

For \dmb, $\gamma_{WD} - \gamma_{M}$ is 26.20$\pm$0.78\kmsec. For a
mass for the M$_{\rm M}$ = 0.184 - 0.185 \msun\ or between 0.192 -
0.193 \msun, the values for i, ii, iii, and iv are 0.49, 0.04, 0.12
and $-$0.04 \kmsec, giving 26.82 \kmsec\ for the white dwarf redshift.

If a carbon-oxygen core white dwarf with metallicity Z=0 and a
hydrogen envelope of fractional mass 10$^{-4}$ is assumed instead, the
values obtained for the masses of the white dwarf are 0.72 and 0.59
\msun\ for \dma\ and \dmb\ respectively.

\begin{table*}
\caption{New values for the M dwarf parameters depending on its
  filling factor $f$ for \dmb. R$_{\rm MCP}$ is the radius calculated
  from the equation of Caillault \&\ Patterson (1990) given in
  Section~\ref{sec:mass}. A range of values for several parameters are
  given. These are the result of assuming two different masses for the
  outer hydrogen envelope of the white dwarf. See text for details.}
\label{res:wdMd:dist}
\begin{tabular}{llllllll}
f & K$_{\rm M}$ & q & M$_{\rm M}$ & i          & a     & R$_{\rm M}$ &
  R$_{\rm MCP}$\\
  & \kmsec  &   & \msun & $^{\circ}$ & \rsun & \rsun & \rsun \\
\hline
0.0 & 144.72 & 0.30 & 0.184 -- 0.192 & 60 -- 59 & 3.191 -- 3.234 & 0.0
  & 0.239 -- 0.247\\
0.1 & 149.95 & 0.29 & 0.178 -- 0.185 & 63 -- 62 & 3.182 -- 3.225 &
  0.085 -- 0.087 & 0.233 -- 0.240\\
0.2 & 155.58 & 0.28 & 0.172 -- 0.179 & 67 -- 65 & 3.173 -- 3.216 &
  0.171 -- 0.173 & 0.226 -- 0.233\\
0.3 & 161.64 & 0.27 & 0.165 -- 0.172 & 72 -- 70 & 3.165 -- 3.207 &
  0.256 -- 0.260 & 0.219 -- 0.226\\
0.4 & 168.20 & 0.26 & 0.159 -- 0.165 & 80 -- 76 & 3.156 -- 3.198 &
  0.342 -- 0.346 & 0.212 -- 0.219\\
0.5 & 175.30 & 0.25 & 0.152 -- 0.159 & 79 -- 84 & 3.147 -- 3.190 &
  0.427 -- 0.433 & 0.205 -- 0.212\\
0.6 & 183.04 & 0.24 & 0.146 -- 0.152 & 72 -- 74 & 3.138 -- 3.180 &
  0.513 -- 0.520 & 0.199 -- 0.205\\
0.7 & 191.49 & 0.23 & 0.140 -- 0.145 & 67 -- 69 & 3.129 -- 3.171 &
  0.598 -- 0.606 & 0.192 -- 0.198\\
0.8 & 200.76 & 0.22 & 0.133 -- 0.139 & 64 -- 65 & 3.120 -- 3.162 &
  0.684 -- 0.693 & 0.184 -- 0.190\\
0.9 & 210.97 & 0.21 & 0.127 -- 0.132 & 61 -- 62 & 3.111 -- 3.153 &
  0.769 -- 0.780 & 0.177 -- 0.183\\
1.0 & 222.28 & 0.20 & 0.120 -- 0.125 & 59 -- 59 & 3.102 -- 3.144 &
  0.855 -- 0.866 & 0.170 -- 0.176\\
\end{tabular}
\end{table*}

The masses obtained for both white dwarfs are unusually high
indicating that the star probably reached the asymptotic giant branch
(AGB) in its evolution. The initial binaries must have been very wide
in order for this to happen. We notice that the masses calculated here
differ significantly from those given in Section.~\ref{dis}, measured
by fitting the line profiles to stellar atmosphere models. This
discrepancy suggests that the redshift measurements may not be
reliable, perhaps not surprising given the difficulty of separating
the M star emission from the white dwarf absorption. Measurements of
the white dwarf at UV wavelengths would be helpful.

As mentioned in Section~\ref{mdcomp}, the flux modulation seen in
Fig.~\ref{res:Mdflux} for \dmb\ indicates that the M dwarf companion
is strongly irradiated and therefore the value measured for K$_{\rm
M}$ is probably a lower limit for the true radial velocity
semiamplitude. This implies that the M dwarf mass and inclination
given in Table~\ref{res:wdMd:mass} are upper and lower limits
respectively. To calculate how distorted the M dwarf is, we calculate
how much the radial velocity semiamplitude of the companion changes as
a function of its radius and how its mass and inclination are
affected. We present the results in Table~\ref{res:wdMd:dist}. The
radius of the M dwarf is given in terms of a linear filling fraction,
$f$, defined as the ratio of the stellar radius measured from the
centre of mass to the inner Lagrangian point. A value of $f$ = 1
implies that the M dwarf fills its Roche lobe. For these calculations
we have taken M$_{\rm WD}$ = 0.61 and 0.64 \msun\ (the maximum and
minimum values calculated above) and K$_{\rm WD}$ = $-$43.81
\kmsec. If the M star does not deviate too far from the main sequence
we expect it to fill at least 0.4 of its Roche lobe which translates
into a true radial velocity semiamplitude in the range 168 $<$ K$_{\rm
M}$ $<$ 222 \kmsec, and a mass between 0.120 $<$ M$_{\rm M}$ $<$ 0.165
\msun.

\subsubsection{The masses of the M dwarf companions}
\label{sec:mass2}

An independent estimate of the masses of the M type companions can be
done from their absolute infrared magnitudes. Although the white
dwarfs dominate the flux in the optical range in both systems, they
produce only a minor fraction of the infrared luminosity. The
distances of the systems were computed from the parameters of the
white dwarfs given in Section~\ref{dis}. Since the flux contribution
of the M dwarfs in the blue part of the spectra, used for the model
atmosphere fits, is very small, we do not expect systematic effects
caused by spectral contamination.

\begin{table*}
\caption{Infrared properties and mass estimates for the M dwarf companions.}
\label{res:irmass}
\begin{tabular}{lccccc}
& dist (pc) & M$_{\rm J}$ & M$_{\rm H}$ & M$_{\rm K}$ & M/$\msun$\\
\hline
\dma & 34.8$\pm$2.5 & 8.71$\pm$0.16 & 8.18$\pm$0.16 & 7.85$\pm$0.16 &
0.169$\pm$0.010\\
\dmb & 115$\pm$9 & 9.28$\pm$0.18 & 8.80$\pm$0.18 & 8.39$\pm$0.18 &
0.136$\pm$0.009\\
\end{tabular}
\end{table*}

J, H, and K magnitudes were retrieved from the 2MASS point source
catalogue. We computed the white dwarf's contribution using the colour
calibration of Bergeron et al. \shortcite{bwb95} and subtracted it
from the observed fluxes. Corrections are small for \dma\ and do not
exceed 25\% (10\%) for the J (K) band flux of \dmb. Finally, the M
dwarf masses were computed from the calibration of Henry \&\ McCarthy
\shortcite{hm93}. Results are listed in Table~\ref{res:irmass}. Our
error estimates include photometric errors and distance uncertainties
resulting from the spectral analysis of the white dwarf. We adopted
Napiwotzki, Green \&\ Saffer's \shortcite{ngs99} estimates for the
external fit errors.

For both systems, the value obtained for the mass of the M dwarf
companion from its infrared magnitudes is in agreement with that
obtained from the spectral analysis in section~\ref{sec:mass}. In the
case of \dmb, the revised mass after taking into account heating by
the white dwarf is considered.

\section{Discussion}
\label{dis}

Table~\ref{dis:orbits} gives a list of all the detached white-dwarf
binaries with known orbits known up to now. Most values have been
taken from Ritter \&\ Kolb \shortcite{rk03}. The 6 systems discussed
in this paper are also included.

\begin{table*}
\caption{List of all the detached white dwarf binaries with known
  orbital periods (given in days). The type of binary is also given
  where WD = white dwarf; M = M dwarf; sdO/sdB = O/B subdwarf; ?  =
  uncertain. $*$ indicates periods measured in this paper. References
  for the orbital periods not measured in this paper are (a)
  Bragaglia, Greggio \&\ Renzini 1990, (b) Koen, Orosz \&\ Wade 1998,
  (c) Orosz \&\ Wade 1999, (d) Maxted et al. 2000a, (e) Morales-Rueda
  et al. 2003a, (f) Maxted et al.  2000b, (g) Marsh 1995, (h) Marsh et
  al. 1995, (i) Moran et al. 1999, (j) Saffer, Livio \&\ Yungelson
  1998, (k) Napiwotzki et al. 2002, (m) Maxted, Marsh \&\ Moran 2002,
  (n) Holberg et al.  1995, (o) Drechsel et al. 2001, (p) Kilkenny et
  al. 1998, (q) Maxted et al. 1998, (r) Wood \&\ Saffer 1999, (s)
  Orosz et al. 1999 (t) Bruch \&\ Diaz 1998, (u) Gizis 1998, (v) Wood,
  Harmer \&\ Lockley 1999 (w) Delfosse et al. 1999 (x) Napiwotzki et
  al. 2001, (y) Maxted et al.  2000c, (z) Saffer, Liebert \&\
  Olszewski 1988, (aa) Heber et al. 2003, (ab) Karl et al. 2003, (ac)
  O'Donoghue et al. 2003, (ad) Maxted et al. 2002, (ae) Hillwig et
  al. 2002, (af) Maxted et al. 2004, (ag) Kawka et al. 2000, (ah)
  Kawka et al. 2002, (ai) Saffer et al. 1993, (aj) O'Brien, Bond \&\
  Sion 2001, (ak) Sing et al. 2004, (al) Moran, Marsh \&\ Bragaglia
  1997, (am) Morales-Rueda et al. 2003b, (an) Edelmann, Heber \&\
  Napiwotzki 2002, (ao) Napiwotzki et al. 2004, (ap) Fuhrmeister \&\
  Schmitt 2003, (aq) O'Toole, Heber \&\ Benjamin 2004, (ar) Robb \&\
  Greimel 1997, (as) Heber et al. 2004, (at) Raymond et al. 2003, (au)
  G\"{a}nsicke et al. 2004, (av) Wood, Robinson \&\ Zhang 1995, (aw)
  Bruch, Vaz \&\ Diaz 2001, (ax) Pigulski \&\ Michalska 2002, (ay)
  Shimansky, Borisov \&\ Shimanskaya 2003, (az) Rauch \&\ Werner 2003,
  (ba) Chen et al. 1995, (bb) Green, Richstone \&\ Schmidt 1978, (bc)
  Landolt \&\ Drilling 1986, (bd) Bell, Pollacco \&\ Hilditch 1994,
  (be) Pollacco \&\ Bell 1994, (bf) Lanning \&\ Pesch 1981, (bg)
  Bleach et al. 2002, (bh) Vennes \&\ Thorstensen 1994.}
\label{dis:orbits}
\begin{center}
\begin{tabular}{lllllllllll}
WD, sdOB + WD &       &        & & WD, sdOB + M & & & & sdOB + ? & & \\
Object&P$_{\rm orb}$&Type&Ref.& Object&P$_{\rm orb}$& Type & Ref. & Object&P$_{\rm orb}$ & Ref.\\
\hline
WD0957$-$666   & 0.061 & WD/WD  & a, al & 
PG1017$-$086   & 0.073 & sdB/M  & m &
HE0532$-$4503  & 0.266 & ao \\

KPD0422+5421 & 0.090 & sdB/WD & b, c& 
HS0705+6700  & 0.096 & WD/M   & o   &
PG1528+104   & 0.331 & am \\

KPD1930+2752 & 0.095 & sdB/WD & d & 
PG1336$-$018   & 0.101 & sdB/M  & p &
KPD1946+4340 & 0.404 & e \\

PG1043+760   & 0.120 & sdB/WD & e &
GD448        & 0.103 & WD/M   & q &
HE0929$-$0424  & 0.440 & ao \\

PG1101+364   & 0.145 & WD/WD  & g &
MT Ser       & 0.113 & sdO/M  & aw  &
HE1318$-$2111  & 0.487 & ao \\

WD1704+481  & 0.145  & WD/WD  & f & 
HW Vir       & 0.117 & sdB/M  & r &
PG1743+477   & 0.515 & e \\

WD2331+290   & 0.166 & WD/WD  & h & 
HS2237+8154  & 0.124 & WD/M & au &
PG0001+275   & 0.528 & an \\

PG1432+159   & 0.225 & sdB/WD  & i & 
NN Ser       & 0.130 & WD/M   &  ax &
PG1519+640   & 0.539 & am \\

PG2345+318   & 0.241 & sdB/WD & i & 
EC13471$-$1258 & 0.151 & WD/M   & ac &
HE1059$-$2735  & 0.556 & ao \\

HE2209$-$1444  & 0.277 & WD/WD  & ab &
J1129+6637   & 0.171 & WD/M & at &
PG1725+252   & 0.601 & e \\

PG1101+249   & 0.354 & sdB/WD  & j, i & 
HS2333+3927  & 0.172 & sdB/M & as &
PG1247+554   & 0.603 & d \\

Feige 48     & 0.376 & sdB/WD & aq &
GD245        & 0.174 & WD/M   & ay  &
HD188112     & 0.607 & aa \\

HE1414$-$0848  & 0.518 & WD/WD  & k &
BPM71214     & 0.202 & WD/M   & ah &
PG1627+017   & 0.829 & e \\

PG0101+039   & 0.570 & sdB/WD  & i & 
PG1329+159   & 0.250 & sdB/M  & e &
PG1230+052   & 0.837 & am \\

PG1248+164   & 0.732 & sdB/WD & e &
PG1224+309   & 0.259 & WD/M   & s &
HE2135$-$3749  & 0.924 & ao \\

PG0849+319   & 0.745 & sdB/WD & e & 
AA Dor       & 0.261 & sdO/M  & az &
PG0133+144   & 1.238 & e, an \\

PG1116+301   & 0.856 & sdB/WD & e & 
WD2154+408   & 0.268 & WD/M   & ae &
PG1512+244   & 1.270 & e \\

PG0918+029   & 0.877 & sdB/WD & e & 
CC Cet       & 0.287 & WD/M   & ai&
UVO1735+22   & 1.278 & an \\

WD1713+332   & 1.123 & WD/WD  & h & 
RR Cae       & 0.304 & WD/M   & t &
HE2150$-$0238  & 1.322 & ao \\

WD1428+373  & 1.143 & WD/WD  & $*$ & 
TW Crv       & 0.328 & sdO/M  & ba &
KPD2040+3955 & 1.483 & am \\

WD1022+050  & 1.157 & WD/WD  & $*$ & 
WD1042$-$690  & 0.336 & WD/M   & $*$, ag&
HD171858     & 1.529 & e \\

HE1047$-$0436  & 1.213 & sdB/WD & x & 
GK Vir       & 0.344 & WD/M   & bb &
PG1716+426   & 1.777 & e \\

WD0136+768   & 1.407 & WD/WD  & m & 
KV Vel       & 0.357 & WD/M   & bc &
PG1300+279   & 2.259 & e \\

Feige55      & 1.493 & WD/WD  & n & 
RXJ1326+4532& 0.364 & WD/M   & ar &
KPD0025+5402 & 3.571 & e \\

L870-2       & 1.556 & WD/WD  & z & 
UU Sge       & 0.465 & WD/M   & bd &
PG0934+186   & 4.05  & am \\

WD1204+450   & 1.603 & WD/WD  & m & 
V447 Lyr     & 0.472 & sdO/M  & be &
PG0839+399   & 5.622 & e \\

PG1538+269   & 2.50  & sdB/WD  & j & 
V1513 Cyg      & 0.497 & WD/M   & u &
PG1244+113   & 5.752 & am \\

WD1241$-$010   & 3.347 & WD/WD  & h & 
V471 Tau     & 0.521 & WD/K   & aj &
HE115$-$0631   & 5.87  & ao \\

WD1317+453   & 4.872 & WD/WD  & h & 
RXJ2130+4710& 0.521& WD/M   & af&
PG0907+123   & 6.116 & e \\

WD2032+188  & 5.084 & WD/WD  & $*$ & 
HZ 9         & 0.564 & WD/M   & bf &
PG1032+406   & 6.779 & e \\

WD1824+040  & 6.266 & WD/WD  & $*$ & 
PG1026+002   & 0.597 & WD/M   & v &
WD0048$-$202   & 7.45  & ao \\

PG1115+166   & 30.09 & WD/WD  & ad & 
EG UMa       & 0.668 & WD/M   & bg &
WD0940+068   & 8.33  & d \\

             &       &        &    & 
REJ2013+400  & 0.706 & WD/M   & v &
PG1110+294   & 9.415 & e \\

             &       &        &    &
WD2009+622  & 0.741 & WD/M   & $*$ &
PG1619+522   & 15.357 & e \\

             &       &        &    & 
REJ1016$-$0520 & 0.789 & WD/M   & v &
PG0850+170   & 27.81 & e \\

             &       &        &  &
HS1136+6646  & 0.836 & WD/K   & ak \\

             &       &        &  &
IN CMa       & 1.263 & WD/M   & ah \\

             &       &        &  &
BE UMa       & 2.291 & WD/K   & av \\

             &       &        &  &
REJ1629+780  & 2.89  & WD/M   & ap \\

             &       &        &  &
Feige 24     & 4.232 & WD/M   & bh \\

             &       &        &  &
G203$-$047ab   & 14.71 & WD/M   & w\\

\end{tabular}
\end{center}
\end{table*}

Liebert, Bergeron \&\ Holberg \shortcite{lbh04}, Bergeron et al.
  \shortcite{bsl92} and Bragaglia et al. \shortcite{brb95} obtain the
  temperatures and gravities for the systems studied in this
  paper. Their values, together with our determination for the masses
  of the white dwarf, are given in Table.~\ref{dis:tgm}.

\begin{table}
\caption{Temperatures and gravities measured for the bright component
  of the system by fitting the hydrogen line profiles to stellar
  atmosphere models. (a) Bragaglia et al.  \protect\shortcite{brb95},
  (b) Liebert et al. \protect\shortcite{lbh04} and (c) Bergeron et al.
  \protect\shortcite{bsl92}. $V$ represents the $V$ magnitude taken
  from the literature and $M_V$ is the absolute magnitude. The masses
  given have been determined using cooling tracks by Althaus \&\
  Benvenuto \shortcite{ab97}.}
\label{dis:tgm}
\begin{center}
\begin{tabular}{lllllll}
WD & $V$ & T$_{\rm eff} (K)$ & $\log g$ & $M/M_{\odot}$ & $M_V$ &
Ref \\
\hline
1022+050 & 14.18 & 14481 & 7.48 & 0.389 & 10.68 & a\\
1428+373 & 15.40 & 14010 & 7.36  & 0.348  & 10.35 & b\\
1824+040 & 13.90 & 14795 & 7.61 & 0.428 & 10.83 & a\\
2032+188 & 15.34 & 18540 & 7.48  & 0.406  & 10.24 & c\\
1042$-$690 & 13.09 & 21380 & 7.86 & 0.551 & 10.52 & a\\
2009+622 & 15.15 & 25870 & 7.70  & 0.489  & 9.93  & c\\

\end{tabular}
\end{center}
\end{table}

\begin{figure*}
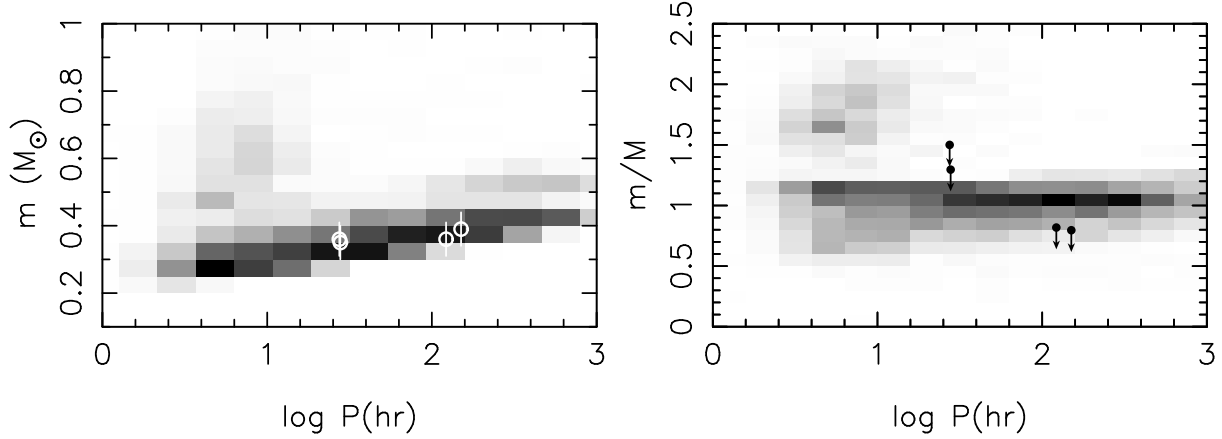

\begin{picture}(100,0)(0,0)
\put(0,0){\includegraphics{P_m_new.ps}}
\put(0,0){\includegraphics{P_q_new.ps}}
\noindent
\end{picture}
\vspace{60mm}
\caption{Left panel: mass of the bright white dwarf for the four DDs
  discussed in this paper as measured by Bragaglia et
  al. \protect\shortcite{brb95}, Liebert et al.
  al. \protect\shortcite{lbh04} and Bergeron et
  al. \protect\shortcite{bsl92} (see Table~\ref{res:mass}) versus
  period distribution. In the grey scale we plot the mass to period
  distribution of DDs according to theory (for the model described in
  Nelemans et al. \protect\shortcite{nyp04}). Right panel: mass ratio
  (only upper limits taken from Table~\ref{res:mass}) for the four DDs
  studied versus period distribution. The theoretical distribution
  according to the Nelemans et al. \protect\shortcite{nyp04} is also
  plotted.}
\label{dis:mpqp}
\end{figure*}

Fig.~\ref{dis:mpqp} presents the theoretical mass versus orbital
period distribution for DDs from the population model as described in
Nelemans et al. \shortcite{nyp04}. The four DDs discussed in this
paper are plotted over the theoretical distribution and fall right in
the expected range of values according to Nelemans et
al. \shortcite{nyp04}.

The direct progenitors of the double white dwarf binaries (giant plus
white dwarf) could have had a wide variety of masses and periods,
leading to inferred efficiencies of the CE that are poorly constrained
(see Nelemans \&\ Tout 2004, fig. 5).

We calculated the possible progenitor systems of the two white dwarf
plus M star binaries. If the high masses inferred from the
gravitational redshifts were right, this means the direct progenitors
of the white dwarfs must have been highly evolved giants. Using the
equations in Hurley, Pols \&\ Tout \shortcite{hpt00} in the same way
as described in Nelemans \&\ Tout \shortcite{nt04} we calculated the
possible progenitor systems. For \dma\ we find possible progenitors
with masses typically in the range 2 -- 3.5 \msun, while for \dmb\ the
progenitors typically have masses between 1.25 -- 3 \msun.  Because of
the rather extreme mass ratios and the large radii of the giants, the
CE is most likely caused by tidal interaction, rather than Roche-lobe
overflow \cite{h80} and the binaries will generally not be
synchronised. We use the formalism derived for star -- planet
interactions by Soker \shortcite{s96} to calculate the separation
between the two stars at which the CE sets in. The required CE
efficiencies are between 0.1 and 1 for \dma\ and between 0.1 and 1.6
for \dmb. Both systems can also be explained with the gamma-algorithm
\cite{nt04}, with values of $\gamma$ around 1.5. If, on the other
hand, the lower white dwarf masses presented in Table~\ref{dis:tgm}
are right, the progenitor masses inferred are in the range 0.9 -- 2.5
\msun\ for \dma\ and 0.75 -- 2 \msun\ for \dmb. The required CE
efficiencies derived in this case lie in the same ranges as for the
more massive white dwarf alternative presented above.

Using the binary results presented in Tables~\ref{res:rv:tab2} and
\ref{res:wdMd:mass}, and the equations from Schreiber \&\ G\"{a}nsicke
\shortcite{sg03}, we have calculated the white dwarf cooling age using
models by Wood, Robinson \&\ Zhang \shortcite{wrz95}, t$_{\rm cool}$,
the orbital period of the binary at the end of the CE phase, P$_{\rm
CE}$, the time it will take for the binary to start mass transfer (and
become a CV) assuming classical magnetic braking (CMB) and assuming
reduced magnetic braking (RMB), t$_{\rm sd}$, and the orbital period
when mass transfer starts, P$_{\rm sd}$. These values are presented in
the top three rows of Table~\ref{dis:sg}. The input masses used for
the white dwarfs are those obtained assuming that they are
carbon-oxygen core white dwarfs. The input masses for the companions
are 0.17\msun\ for \dma\ and 0.120 $<$ M$_{M}$ $<$ 0.165 \msun\ for
\dmb\ (as calculated in Section~\ref{sec:mass}). The value of P$_{\rm
CE}$ for both binaries is very close to their present orbital periods
which indicates that they are very young PCEBs. These systems will not
become CVs within a Hubble time (except perhaps \dma\ if RMB takes
place), assuming $\tau_0$ = 1.3$\times$10$^{10}$ yrs \cite{fms01}, and
when they do, their orbital periods will place them below the CV
period gap.

The evolutionary properties of \dma\ and \dmb\ are also calculated by
using the lower white dwarf masses presented in Table~\ref{dis:tgm}
and the masses of the M dwarfs obtained in
Section~\ref{sec:mass2}. These are presented in the bottom two rows of
Table~\ref{dis:sg}. The values for all parameters in this case are
similar to those obtained for the larger white dwarf masses and the
conclusions reached are equivalent.

\begin{table}
\caption{Evolutionary properties of \dma\ and \dmb. The top three rows
  present the results obtained assuming the white dwarf masses given
  in Table~\ref{res:wdMd:mass} (0.72 and 0.59 $\msun$
  respectively). $^1$ and $^2$ indicate calculations obtained for
  \dmb\ by assuming M$_{\rm M}$ = 0.12 and 0.165 \msun\
  respectively. The bottom two rows present the results obtained when
  the lower white dwarf masses given in Table~\ref{dis:tgm} are
  assumed instead. In this case the M dwarf masses calculated in
  Section~\ref{sec:mass2} are used. The values of $t_{\rm cool}$ and
  $t_{\rm sd}$ given in the table are in fact the logs of $t_{\rm
  cool}$ and $t_{\rm sd}$ in years.}
\label{dis:sg}
\begin{center}
\begin{tabular}{lllllll}
WD & $t_{\rm cool}$ & \multicolumn{2}{c}{P$_{\rm CE}$ (d)} &
  \multicolumn{2}{c}{$t_{\rm sd}$} & P$_{\rm sd}$\\

 & & CMB & RMB & CMB & RMB & (d)\\
\hline
1042$-$690 & 7.92 & 0.3381 & 0.3375 & 10.30 & 10.08 & 0.073 \\
2009+622$^1$ & 7.31 & 0.7412 & 0.7411 & 11.43 & 10.96 & 0.055 \\
2009+622$^2$ & 7.31 & 0.7412 & 0.7411 & 11.30 & 10.78 & 0.071 \\
\hline
1042$-$690 & 7.65 & 0.3376 & 0.3372 & 10.39 & 10.09 & 0.073\\
2009+622   & 7.20 & 0.7412 & 0.7410 & 11.43 & 10.87 & 0.061\\
\end{tabular}
\end{center}
\end{table}

\section{Conclusions}

We have obtained the orbital solution for four DD systems and two
white dwarf - M dwarf binaries. We find that the white dwarf
companions for the four DDs studied contribute between 10 and 20 per
cent of the total luminosity and remain undetected. The masses and
periods obtained for these systems agree with theoretical mass period
distributions (based on Nelemans et al. \shortcite{nyp04}).

In the case of the white dwarf - M dwarf binaries we have been able to
measure the motion of both components. We find that there are
signatures of strong irradiation of the surface of the M dwarf
component in \dmb. The white dwarf masses calculated from their
gravitational redshift are unusually high compared to those measured
by fitting their hydrogen lines with stellar atmosphere models, and
could be the result of the evolution of giant stars with masses
between 1.25 and 3.5 \msun\ that went through a CE phase as a result
of tidal interaction. These two binaries are young PCEBs that will
evolve into CVs, although not within a Hubble time, with orbital
periods below the period gap.

\section*{Acknowledgements}

LMR was supported by a PPARC post-doctoral grant and by NWO-VIDI grant
639.042.201 to P.J. Groot during the course of this research. GN is
supported by PPARC and an NWO-VENI grant. TRM acknowledges the support
of a PPARC Senior Research Fellowship. RN acknowledges support by a
PPARC Advanced Fellowship. The authors would like to thank
M. R. Schreiber for his help with implementing the calculations for
this paper. The reduction and analysis of the data were carried out on
the Southampton node of the STARLINK network. We thank PATT for their
support of this program.

\label{lastpage}

\end{document}